\documentclass{article}

\usepackage{PRIMEarxiv}

\usepackage[utf8]{inputenc} 
\usepackage[T1]{fontenc}    
\usepackage{hyperref}       
\usepackage{url}            
\usepackage{booktabs}       
\usepackage{amsfonts}       
\usepackage{nicefrac}       
\usepackage{microtype}      
\usepackage{lipsum}
\usepackage{fancyhdr}       
\usepackage{graphicx}       
\graphicspath{{media/}}     

\usepackage{tikz}
    \usetikzlibrary{arrows,positioning,shapes}
\usepackage{pdflscape}
\usepackage{ragged2e}     
\usepackage{xltabular}    
\newcolumntype{P}[1]{>{\RaggedRight\arraybackslash}p{#1}}
\newcolumntype{Y}{>{\RaggedRight\arraybackslash}X}
\usepackage[caption=false]{subfig}

\title{The Art of Storytelling in Authoritarian Regimes: Crafting State Narratives on Chinese Social Media
\thanks{\textit{\underline{Corresponding author}}: 
\textbf{Dr. Yan Wang, yan.wang-5@manchester.ac.uk.}} 
}

\author{
  Dr. Ting Luo \\
  Associate Professor in Government and AI \\
  Department of Public Administration and Policy \\
  University of Birmingham \\
  United Kingdom \\
  \texttt{t.luo.1@bham.ac.uk}
   \And
  Dr. Yan Wang \\
  Assistant Professor in Social Statistics \\
  Department of Social Statistics \\
  University of Manchester \\
  United Kingdom \\
  \texttt{yan.wang-5@manchester.ac.uk} \\
}

\begin{document}
\maketitle

\begin{abstract}
This article examines how authoritarian regimes construct state narratives about politically consequential events. Building on the narrative policy framework and existing research on authoritarian propaganda, we propose two dimensions that shape narrative construction: legitimacy implications -- whether events enhance or threaten regime legitimacy, and citizen verification capacity -- the extent to which citizens can evaluate official narratives through alternative sources. Using quantitative narrative analysis of Chinese social media posts by government, state media, and celebrity accounts, we extract subject–verb–object (SVO) triplets to map dominant narrative structures across four major events. Our findings show that legitimacy implications of the event shape regime's efforts in storytelling and the beliefs highlighted in the narratives, while citizen's verification capacity could balance the strategic choice between a top-down manipulation and bottom-up responsiveness of state narratives. Together, the results reveal propaganda as a complex process of narrative construction adaptive to specific contexts, offering new insights into how dynamic storytelling sustains authoritarian resilience.
\end{abstract}


\section{Introduction}
Contemporary authoritarian regimes, particularly tech savvy ones like China, have evolved beyond traditional censorship and repression toward more sophisticated approaches \cite{deibert2010access, munger2019elites}. The Chinese state conveys messages through official media that have become more subtle, sophisticated, engaging, and credible \cite{stockmann2013media}, a form of “soft propaganda'' that has proven more persuasive than traditional hard propaganda \cite{huang2018pathology,MattinglyYao}. The key to this soft propaganda lies in the use of a comprehensive framework that connects past, present, and future and a causal interpretation of all punctual events, making it cognitively costly for individuals to evaluate each claim independently \cite{przeworski2023formal,horz2017strategic}. Such frameworks concern interpretations, rather than facts \cite{przeworski2023formal}. 

While scholars debate whether propaganda operates through top-down manipulation or by affirming citizens' existing beliefs and emotions, both perspectives recognize that regimes must craft and disseminate compelling stories to shape political attitudes \cite{greene2019putin, shirikov2024rethinking,matovski2021popular,tang2016populist, su2016selective,ding2022performative,greene2022affect}. Yet existing research has paid relatively less attention to the interpretive frameworks that build and promote state narratives. What stories do authoritarian regimes craft? In this article, we focus on narratives \cite{miskimmon2014strategic} as the interpretative framework of propaganda and investigate: How do authoritarian regimes construct state narratives on political and social events? What are the specific storylines? How do their communication strategies and underlying beliefs vary? Our analysis focuses on the production side of propaganda -- how the state craft and coordinate narratives, rather than examining citizen reception. 

Building on the “narrative policy framework'' \cite{jones2014science, boscarino2022constructing, gupta2018advocacy, merry2019angels} as an analytical tool, we examine how authoritarian regimes craft interpretations of reality through the selective appropriation of actors and actions, the strategic portrayal of relationships, and the construction of underlying beliefs. We focus on politically consequential events that vary along legitimacy implications -- whether events are beneficial or threatening to the regime, and citizen verification capacity -- whether citizens can access alternative sources to evaluate official narratives. Using quantitative narrative analysis that extracts semantic triplets -- subject-verb-object (SVO) -- from text data, complemented by close readings of dominant triplets and representative posts, we analyze social media content produced by government, media and celebrity users in China's cyberspace and uncover the art of storytelling. Our findings show that the legitimacy implications of events shape the beliefs embedded in the narratives, while the citizen verification capacity constrains explicit strategies of the narratives. When events create opportunities to improve legitimacy, narratives tend to channel national pride and achievement toward regime credit, cultivating a belief in a nationalist China whose success is attainable only under the current government’s leadership. By contrast, when events potentially challenge the legitimacy of the regime, state narratives pivot to emphasize government responsiveness and problem-solving capacity, fostering a belief in benevolent and competent leadership. Meanwhile, citizen verification capacity constrains narrative strategies. When citizens can verify state claims through lived experience or accessible evidence, regimes adopt more bottom-up, responsive storytelling that acknowledges citizen-verified realities and emphasize mobilization and government actions; when citizens lack verification means, regimes possess greater latitude for top-down manipulation and selective framing.

This article offers theoretical and methodological contributions to the study of authoritarian politics. Theoretically, this study bridges ongoing debates on propaganda in authoritarian regimes by uniting two perspectives often treated separately: propaganda as top-down manipulation versus propaganda as the affirmation of existing beliefs and emotions of citizens. We move beyond a static view of propaganda as mere control and conceptualize it as a form of adaptive narrative construction. This perspective highlights that propaganda is not uniform but varies according to context: under some conditions, events with low citizen verification capacity, it functions as top-down manipulation, while under others, events with high citizen verification capacity, it becomes more bottom-up and responsive, engaging with citizen sentiments and experienced realities. By emphasizing these dynamics, our study offers a more nuanced and dynamic understanding of propaganda in authoritarian politics.

We also make a methodological contribution by employing quantitative narrative analysis through SVO extraction, adapting a well-established technique to the context of Chinese-language texts. Rather than seeking to automate interpretation, this approach serves as an exploratory tool that helps identify dominant state narratives from large volumes of text data. The extracted SVO structures provide a systematic overview of recurring actors, actions, and relationships, allowing us to zoom in on dominant narrative patterns for closer qualitative examination. This combined computational–interpretive strategy enables a more grounded understanding of how propaganda operates and evolves in contemporary digital authoritarianism.

\section {Theoretical Framework: Understanding State Narrative Construction}

To understand how authoritarian regimes construct state narratives, we draw on theoretical insights about narratives as tools of political communication with the public. Narratives are essential stories through which people translate knowing into telling and represent social reality through structures of meaning \cite{patterson1998narrative,white1980value}. The power of narratives as tools for manipulation, persuasion, and connection stems from their inherent subjectivity: narratives do not always represent what is real, but rather the narrators' perspective on what is canonical or real \cite{monroe1998heart}. Through selective appropriation of elements and strategic arrangement of the sequence and relationships of elements within a plot, narrators construct particular interpretations of reality \cite{somers1994reclaiming}. Thus, narratives are not mere reports of reality; instead, they reveal how people cognitively weave together disparate facts to make sense of social or political reality, which in turn shapes political behavior \cite{patterson1998narrative}. 

The Narrative Policy Framework (NPF) offers useful analytical scheme for examining political narratives \cite{jones2010narrative, jones2014science}. The NPF identifies three key components of policy narratives: narrative elements, narrative strategies, and policy beliefs. Narrative elements are the distinctive narrative structures of a story that buttress the policy preference, including the structural components like settings, context, plots, characters, and the moral of the story. Narrative strategies are the tactical portrayal and use of narrative elements to expand, contain, or otherwise manipulate involvement in the policy arena. Strategies commonly seen in the policy-making process include “devil shift'' \cite{shanahan2011policy}, change of scope of conflict, or using focusing events and issue expansions \cite{merry2016constructing}. Policy beliefs can be thought of as a set of values and beliefs that critically link the “is’s'' and “ought’s'' of the world to form coalitional actors’ interpretations of reality \cite{jones2014science}. Structurally, narratives can be understood as sequences of actors with agency pursing goals through their actions \cite{franzosi2012ways}. This understanding provides the theoretical foundation for analyzing narrative structures through SVO relationships, where subjects, verbs and objects map to NPF's narrative elements, and the patterns across SVO relationships reveal narrative strategies and beliefs. The selective appropriation of actors and actions, the strategic portrayal of relationships, and the underlying beliefs these choices reveal all offer insights into how authoritarian regimes craft interpretations of reality.

Our objective of understanding the state narratives also fits well in the literature of propaganda in authoritarian regimes. Existing work has extensively examined explicit propaganda strategies, from censorship and information manipulation to attention diversion and selective attribution \cite{king2013censorship, king2017chinese,munger2019elites, rozenas2019autocrats}. However, these studies quite often focus on particular cases, especially the role of propaganda during crises and on controversial issues \cite{king2017chinese, munger2019elites}, assuming events as simply blank canvases for deploying propaganda techniques, with less attention to potential systematic patterns across diverse event types. Our approach, instead, examine whether the properties of events themselves might create different possibilities and constraints for state narrative construction. Essentially, the regimes must allocate limited attention and resources to events that carry implications for their rule, whether opportunities to build support or threats requiring damage control. 

Two factors surface in existing studies as potentially important constraints underpinning political and social events. First, the literature's focus on the role of propaganda during crisis or protests highlights an important factor: legitimacy implications, referring to whether events are beneficial or threatening to the regime. Some events are inherently regime-bolstering, such as ceremonial events or the hosting of major international events. Some can threaten regime legitimacy and often emerge unexpectedly. The 2011 Wenzhou high-speed rail crash in China exemplifies this challenge, where the state had to manage a sudden crisis that claimed the lives of many and sparked intense online discussion and public outcry \cite{gorman2021managing}. A substantial body of policy studies research examines such events -- often referred to as “focusing events'' -- which are sudden occurrences with concentrated harms in one place and time that become known to policymakers and the public simultaneously \cite{birkland1998focusing, birkland2013defining, kingdon2003agendas}. These crises or disasters often bring latent issues to the policy agenda.

Second, existing research suggests that the extent to which citizens can verify official narratives against alternative sources affects the degree of flexibility authoritarian regimes have in crafting interpretations. Some events, particularly domestic ones, directly affect citizens’ everyday lives, making it easier for them to evaluate official narratives against multiple observables and alternative information channels. Such information may come from social interactions, various national and local media, involvement in organizations and activities, or direct personal observation \cite{books1988studying}. For example, with economic news, citizens can compare official news reports against observables such as personal income, market prices, or queues at unemployment offices \cite{rozenas2019autocrats}. To understand political boundaries and the degree of repression by the regime, Chinese political activists, particularly journalists and lawyers, often triangulate official narratives with information gathered from their trusted social networks \cite{Amplify2012}. In contrast, other events, particularly some international events involving foreign actors and pertaining to security, sovereignty, or diplomatic issues, present greater challenges for citizens to benchmark official information. Citizens typically lack the same rich information system available for domestic events, face linguistic and cultural barriers to alternative sources, and encounter state-imposed constraints such as China’s Great Firewall, which restricts access to international perspectives. This information asymmetry makes it substantially more difficult for citizens to verify and evaluate official narratives, giving regimes greater latitude in narrative construction. 

These two dimensions, legitimacy implications and citizen verifiability capacity, provide useful analytical lenses for our exploratory analysis. We do not assume these dimensions are exhaustive or deterministic, nor that they operate independently. Indeed, they likely intersect in complex ways: a threatening event that is verifiable poses different narrative challenges than a beneficial event that is unverifiable. Rather than predicting fixed outcomes, we use these dimensions as analytical lenses to explore potential patterns in how event characteristics shape narrative construction in authoritarian regimes.

\section{Constructing a Pro-regime State Narratives on China’s Cyberspace}

In order to reap the benefits of the Internet while also containing its potential destabilizing effects, the Chinese state has continuously upgraded its approach to cyberspace management and information control. The first decades of the 2000s under the Hu administration were marked by considerable space for citizens to express concerns and criticize the authority around social issues, with the state emphasizing the use of online channels for “public opinion supervision'' to strengthen government legitimacy and deter large-scale collective action \cite{cairns2017china, kluver2005us, qiang2011battle}. However, the collision of two high-speed trains in Wenzhou in 2011, coinciding with the Arab Spring, led the government to reconsider its soft and open approach to online public debates, as the government struggled to contain coverage of the incident online and the widespread criticism and public outcry regarding the government’s handling of the incident and corruption behind the construction of the railway \cite{luo2021nine}. As a result, opinion supervision gave way to opinion guidance, a proactive approach greatly reinforced under the current leadership of Xi \cite{cairns2017china, StockmannLuoforthcoming}. 

The current leadership under Xi viewed the Internet as the main battlefield for public opinion, implementing several measures to seize the “commanding heights” of social media \cite{cairns2017china, creemers2017cyber, taylor2022china}.\footnote{Rang Zhuliu Meiti Laolao Zhanlin Chuanbo Zhigaodian (Let mainstream media firmly occupy the commanding heights of communication—Notes on the 12th collective study session of the Central Political Bureau), Xinhua Net, \url{https://china.chinadaily.com.cn/a/201901/26/WS5c4bbf13a31010568bdc6953.html}, Retrieved November 22, 2023.} Beyond reinforced regulations aiming to end anonymity and make it easy for the regime to identify and silence dissent \cite{creemers2015pivot, creemers2017cyber}, the state has intensified its efforts in disseminating official ideologies and discourse, aiming to construct pro-regime narratives and guide public opinion. While paid web commentators, also known as the fifty-cent party, were employed to actively channel and shape online public discussion towards the direction favorable to the regime \cite{bandurski2008china,han2015manufacturing, han2018contesting, king2017chinese, miller2016automated}, the Party’s mouthpieces, such as Xinhua and People’s Daily, were encouraged to set the agenda and lead online public discussion \cite{StockmannLuoforthcoming}. Official microblogging accounts proliferated online with the aim of promoting positive images of the government and guiding public opinion \cite{esarey2015winning}. Weibo and many other social media platforms become the key arena for the state to strategically and actively construct and promote pro-regime narratives. On Weibo, Blue Vs -- usually verified accounts of government officials, public institutions, media outlets, enterprises, and non-governmental organizations -- actively promote pro-regime messages, a phenomenon described by netizens as “Blue V/Sea Operation''.\footnote{Dang Zhuliu Meiti Buxiang Rang Laobaixing Shuohua, Jiu Fadong “Lan V Xingdong”(When mainstream media doesn't want people to speak their minds, they launch the ‘Blue V Operation'), See \url{https://zhuanlan.zhihu.com/p/523559709}, Retrieved November 21, 2023.} For example, during the Shanghai Lockdown, when faced with mounting criticism and public outcry from local residents, Blue Vs inundated Weibo with positive posts and comments about the government and the lockdown. Unlike paid web commentators who fabricate support in the guise of unsolicited comments from citizens, Blue Vs actively set the agenda for public discussion and engage patriotic netizens in creating and disseminating pro-regime narratives.\footnote{Rang Zhuliu Meiti Laolao Zhanlin Chuanbo Zhigaodian (Let mainstream media firmly occupy the commanding heights of communication—Notes on the 12th collective study session of the Central Political Bureau), Xinhua Net, \url{https://china.chinadaily.com.cn/a/201901/26/WS5c4bbf13a31010568bdc6953.html}, Retrieved November 22, 2023.} In this context, government and media accounts construct pro-regime narratives, while celebrity accounts -- an influential group of users that frequently repost and comment on the pro-regime content \cite{wang2023politicizing} -- play a pivotal role in facilitating the construction and dissemination of state narratives on Weibo. 

\section{Research Design}

\subsection {\textit{Design and data collection}}

Our research focuses on the role of influential opinion leaders in constructing state narratives on social media. We choose to focus on Weibo, China's Twitter-like platform, because it has become the key avenue for the Chinese state to exercise opinion guidance. Under the current leadership of Xi, media outlets are tasked with constructing pro-regime narratives and guiding public opinion to advocate the party's positions and viewpoints.\footnote{Yang Liqun, “Zai Buduan Chuanxin Zhong Zhangwo Xinwen Yulun Gongzuo Zhudongquan Zhudaoquan''(Grasping the Initiative and Leadership in News and Public Opinion Work through Continuous Innovation), Hongqi Wengao, 2019, \url{http://www.qstheory.cn/dukan/hqwg/2019-07/10/c_1124734850.htm}, accessed September 24, 2024.} As such, media accounts and government accounts play similar roles in propaganda. Beyond government and media accounts, existing studies have demonstrated that celebrity accounts on Weibo have been co-opted by the state to promote its political objectives \cite{chen2023celebrities,sullivan2019celebrity,schneider2017china}. Therefore, we focus on two main group of organizational accounts -- government/media accounts, and celebrity accounts -- and collect content posted by these accounts. 

We first drew a random sample of 300,000 Weibo accounts from a pool of 20,000,000 active users on Weibo constructed by Hu and his co-authors \cite{HuWeiboCov}. We then collected all posts posted by these accounts during the five months between January 1st to May 5th, 2022. We selected the first five months of the year as our observation period because it encompasses China's “Two Sessions'' -- the annual meetings of both local and national legislature and high-profile advisory bodies. This is a time of heightened policy debate and public attention to social issues. During this period, the convergence of political events and social discussions creates both the necessity and opportunity for intensive state narrative-building, making it an ideal window for capturing state narrative construction efforts. 

From this full dataset, we then filter the two categories of verified accounts -- government/media accounts and celebrity accounts, and keep only posts posted by these types of users, which constitute the dataset for the analysis of this paper.\footnote{Verified accounts status is similar to the blue Verified badge on Twitter/X. More info on verified accounts on Weibo and how to get verified status, please see \url{https://china-digital.com/blogs/weibo-account-registration-and-promotion/}. Celebrity accounts here are commonly referred to as Red Vs and Yellow Vs on Weibo, including verified individual accounts such as stars, video makers, and various bloggers (specializing in education/medicine/sports areas).} This results in 2,959,577 posts by 33,782 verified accounts. Table 2 presents the distribution of accounts and posts in our dataset. 

\begin{table}[htbp]
    \centering
    \caption{Description of selected verified accounts and posts}
    \begin{tabular}{p{1.7cm}|p{1.2cm}|p{1.2cm}|p{1.2cm}|p{0.9cm}}
    \hline \hline 
     & N of accounts & N of posts & Accounts ratio & Post ratio \\\hline
    {Celebrities} & {31112} & {2113421} & {0.82} & {0.71} \\ \hline
    {Government/ \newline Media} & {2670} & {846156} & {0.08} & {0.29} \\\hline 
    Total & 33782 & 2959577 & 100\% & 100\% \\\hline \hline 
    \end{tabular}
\end{table} 

\subsection {\textit{Analytical procedure}}

Because of the exploratory nature of this study, we employ two steps to identify cases of state narrative constructions for our analysis. First, we use topic modeling to systematically identify events that generated a large volume of posts by government/media and celebrity accounts during our study period. High posting volume signals instances where there are substantial efforts by the state in actively constructing and disseminating state narratives. This data-driven approach allows us to capture the breadth of state narrative activity without pre-imposing assumptions about which events matter. Topic modelling provides an ideal technique for identifying such information-rich cases, as it helps reveal major themes from a large volume of text data \cite{nelson2020computational,chen2023what}. We use an unsupervised machine learning model -- the Non-negative Matrix-Factorisation (NMF), a linear-algebraic model that reduces high-dimensional data into low dimensions with non-negative components \cite{berry2005understanding}, to identify patterns in the dataset and group text data into topics. We present the validations of the topic modeling results and key topics in the appendix. 

Topic modeling identified three dominant topics corresponding to major events during the study period: the Shanghai COVID-19 lockdown (March-May 2022), the Beijing Winter Olympics (February 2022), and the Russia-Ukraine war (beginning February 2022). These events generated substantial official communication activity and represent different types of narrative challenges: a domestic crisis management situation (Shanghai lockdown), a prestige event with international visibility (Olympics), and an international conflict with implications for China's geopolitical positioning (Russia-Ukraine war).

Second, we supplement this with manual identification of legitimacy threatening events requiring state's “damage control''. While censorship or attention diversion has been well-documented response to sensitive events perceived as regime-threatening in authoritarian regimes \cite{king2013censorship, king2017chinese}, extensive censorship can undermine media credibility and fundamentally political trust \cite{GEHLBACH2014163,StockmannLuoforthcoming}. While heavily censored events may leave limited digital trace in our data, such events still needs engagement of the state narrative rather than left in blank or purely suppressed. Meanwhile, the potential limitation brought by censorship is manageable considering the nature of our exploration focuses on the observable state narrative in cyberspace. We identify this through manual review of major events and news coverage during the study period. Through manual review, we identified the chained woman incident (late January \& February 2022) in a poor county called Fengxian in eastern China (hereafter, “the Chained Woman Incident'' or “Fengxian Event'') \footnote{More information on the case see reports here \url{https://www.washingtonpost.com/world/2022/02/25/xuzhou-chained-woman-china/} and \url{https://www.nytimes.com/2022/03/01/business/china-chained-woman-social-media.html}.}. This case came to light by chance after celebrity social media accounts exposed the poor condition of a woman who had been kept in chains and subsequently provoked substantial public outcry. The state response has been mixed, involving both selective censorship, attention diversion, and narrative construction. The four selected events are presented in Table 3 along with total number of posts for each event. 

\begin{table}[h!]
\centering
\caption{Number of Social Media Posts on Selected Events}
\begin{tabular}{l|c}
\hline \hline
\textbf{Event} & \textbf{Number of Posts} \\
\hline 
Shanghai COVID-19 Lockdown & 218772 \\ \hline
Beijing Winter Olympics & 149526 \\ \hline
Russia--Ukraine War & 180167 \\ \hline
Chained Woman Incident & 3316 \\ 
\hline \hline
\end{tabular}
\end{table}

After selecting the cases, we adopted a structural approach in quantitative narrative analysis \cite{tilly2008contentious, franzosi2004words, franzosi2010quantitative, franzosi2012ways} to capture narratives embedded in large volumes of text data. Specifically, we use Semantic Role Labeling (SRL), a computational linguistics technique that identifies the core semantic structure of sentences. We extract “semantic triplets'': SVO structures with their modifiers that capture the relational grammar of each statement \cite{franzosi2010quantitative}. For example, the sentence “Ukraine starts the 30-day national emergency'' yields the triplet: Ukraine - starts - national emergency.

We modified the RELATIO package\footnote{Python RELATIO package, more information see \url{github.com/relatio-nlp/relatio}.} developed by \cite{ash2023relatio}, replacing its embedded NLP library with HanLP \cite{he-choi-2021-stem}, which is better suited for Chinese text processing. The HanLP parsing functions allow us to extract triplets that satisfy agent-patient relationships with their respective verbs, creating a structured dataset of actor-action-object relations from the original posts.

Raw triplet extraction produces thousands of variants expressing similar relationships. To identify meaningful narrative patterns and reduce dimensions, we conducted iterative refinement. We first perform in-depth reading of narratives identified from the 10 percent random sample of each event. Based on this reading, we developed synonym dictionaries and identified clustering patterns. We then applied these dictionaries and patterns to the full corpus, grouping similar triplets. At the final step, we visualized the relationships between actors and actions as narrative networks for each event. Nodes represent subjects and objects, while arrows indicate the direction of actions with verbs displayed on the links. Arrow thickness corresponds to the frequency of each SVO triplet, and node size reflects actor prominence. For interpretability, we mapped all narratives for the Chained Woman case given its scarcity of posts, and the top 20 percent of narratives from government/media and celebrity accounts for the remaining three events. The resulting figures, included in the Appendix, provide an overview of the broader narrative structure and relationships among key actors. In the main text, we present the top five SVO triplets for each event in Table 3, along with their frequencies and sample sentences, to facilitate clear reading and interpretation.

To ground our extracted patterns in actual state narratives, we selected representative full posts that illustrate how the state articulates actor-relationships in concrete language. We prioritize posts that contain the most prevalent triplets by the two groups of actors in official communication. The Appendix contains a collection of these sample posts. 

\section{Findings}

We use legitimacy implications and citizen verification capacity as analytical lenses to explore patterns across a spectrum of event characteristics. Rather than forcing cases into rigid categories, we acknowledge that real events are messier and have mixed properties that do not fit neatly into theoretical typologies. The selected four events exhibit marked variation across both dimensions. Legitimacy implications vary substantially across the cases. The Beijing Winter Olympics represents a state-orchestrated showcase event designed to enhance regime legitimacy and demonstrate national achievement. The Russia-Ukraine War created legitimacy tensions for the Chinese government: China's strategic partnership with Russia and its long-standing emphasis on sovereignty and non-interference made it reluctant to condemn the invasion, yet mounting international criticism risked diplomatic isolation and exposed tensions with its non-interference stance. Both the Shanghai Lockdown and Chained Woman event posed threats to regime legitimacy, though in different ways. The Shanghai Lockdown, despite being an official policy consistent with established zero-COVID protocols since 2020, developed into a crisis as implementation failures produced food shortages, forced separations, and widespread chaos that undermined rather than bolstered public support for the regime's pandemic response. The Chained Woman event unexpectedly push a deep-seated problem -- human trafficking -- into public attention after social media influencers exposed the case. 

Citizen verification capacity also varies significantly across the four events. The Shanghai Lockdown sits at the high end of this dimension, as residents could directly benchmark official narratives against their lived experiences of shortages, quarantine conditions, and policy implementation. The Chained Woman incident exhibits more complex and dynamic verification capacity. The initial viral videos provided citizens direct visual evidence that contradicted later official accounts, but subsequent state censorship, content removal, and attention diversion strategies progressively constrained citizens' ability to access alternative information and verify evolving official narratives about the case. The Beijing Olympics represents relatively low citizen verification capacity. Because the event coincided with Covid-19 and the spread of the Omicron variant, ticket sales to general public were canceled with only limited invited spectators admitted by invitation. Citizens rely predominantly on official channels for information regarding the event, with limited alternative sources of information. The Russia-Ukraine War represents even lower verification capacity, because the Great Firewall blocked access to international news sources and alternative perspectives, and geographic distance and language barrier left citizens unable to independently assess competing claims. 

These varying configurations of legitimacy implications and verification capacity created distinct conditions for state narrative construction and contestation, which we examine in detail below. Figure 1 represents the approximate and relative placement of the four cases on the two dimensions. Table 3 shows the top five triplets for each event by the two groups of Weibo accounts. 

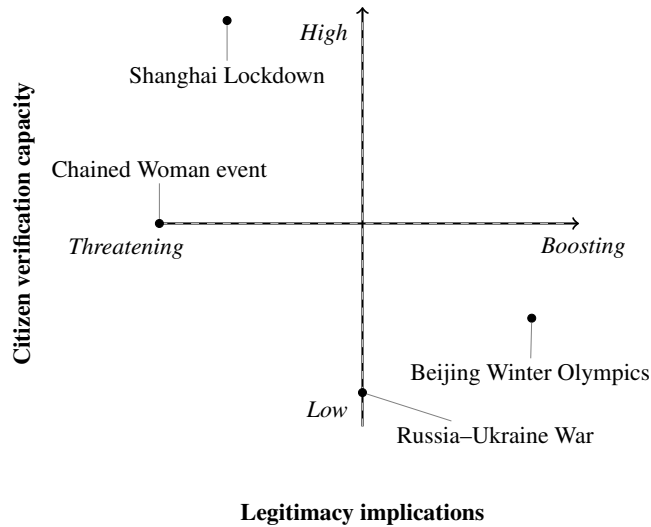
\begin{figure}[htbp]
\centering
\begin{tikzpicture}[scale=0.9, every node/.style={font=\small}]
  \tikzset{
    event/.style={circle, draw, fill=black, minimum size=3pt, inner sep=0pt}, 
    every pin/.style={fill=white, rounded corners=1pt, inner sep=2pt, font=\small},
    pin edge/.style={<-, shorten <=2pt}
  }

  \draw[->, thick] (-3,0) -- (3.2,0);
  \draw[->, thick] (0,-3) -- (0,3.2);

  \node[below] at (0,-4) {\textbf{Legitimacy implications}};
  \node[rotate=90] at (-5,0) {\textbf{Citizen verification capacity}};

  \node[below left,yshift=-2pt]  at (-2.5,0) {\emph{Threatening}};
  \node[below right,yshift=-2pt] at ( 2.5,0) {\emph{Boosting}};

  \node[above left,xshift=-2pt]  at (0, 2.5) {\emph{High}};
  \node[below left,xshift=-2pt]  at (0,-2.5) {\emph{Low}};

  \draw[dashed, gray!50] (-3,0) -- (3,0);
  \draw[dashed, gray!50] (0,-3) -- (0,3);
  \node[event, pin=-92:{Beijing Winter Olympics}] at (2.5, -1.4) {};
  \node[event, pin=-45:{Russia--Ukraine War}] at (0, -2.5) {};
  \node[event, pin=90:{Chained Woman event}] at (-3, 0) {};
  \node[event, pin=-90:{Shanghai Lockdown}] at (-2.0, 3) {};
\end{tikzpicture}
\caption{Analytical positioning of cases along legitimacy implications and citizen verification capacity dimensions}
\end{figure}

We conducted a systematic comparison of narrative construction across all four cases. This comparative exploration reveals three systematic patterns, which we present below. 

\subsection{Finding 1: Legitimacy Implication Shapes Narrative Beliefs}
Across all cases, the legitimacy implications of events -- whether they potentially enhance or threaten regime legitimacy -- shape what regimes wanted citizens to believe. This pattern manifested in two distinct beliefs: belief in national achievement and regime capability when legitimacy could be boosted, and belief in benevolent, responsive governance when legitimacy was threatened.

When events creates opportunities for legitimacy enhancement, narratives tend to channel to national pride and achievement toward regime credit. The Beijing Winter Olympics exemplified this approach. Both government/media accounts and celebrity accounts constructed a consistent narrative dominated by two key actors: Beijing and China. Top triplets establish positive framing emphasizing national competence and global standing. Top triplets like “China is prepared'' and “the world expects/awaits China'' positioned China as organizationally competent and internationally recognized. Celebrity accounts amplified this through commercial nationalism, with the top triplet “China has Anta Sports'' highlighting the Chinese sportswear company as official Olympic partner -- reinforcing broader national initiatives including the “China Dream'' and “Made in China 2025'' plan aimed at establishing China as a global manufacturing superpower.\footnote{“Shixian Zhonghua Minzu Weida Fuxing Zhongguo Meng de Guanjian Yibu” (A Key Step in Realizing the China Dream of Great Rejuvenation of the Chinese Nation), available at \url{http://www.qstheory.cn/zhuanqu/2021-07/22/c_1127683059.htm}, accessed September 25, 2024. “Li Keqiang tan ‘Zhongguo zhizao 2025': Cong Zhizao Daguo Maixiang Zhizao Qiangguo'' (Li Keqiang on “Made in China 2025'': From a Manufacturing Giant to a Manufacturing Power), Chinese Government Website, available at \url{https://www.gov.cn/premier/2017-08/10/content_5216727.htm}, accessed September 25, 2024.}

The linkage between commercial success, national pride, and regime legitimacy emerged vividly in individual posts. A post by an influencer on Febuary 17th celebrated: “At the Beijing Winter Olympics, ANTA seems to be saying: ‘If there's even one camera angle without our logo spotted, I’ll be heartbroken, OK?' Gotta admit -- Eileen Gu’s golden dragon suit is super stylish, keeps her warm and makes her look slim. And the ink-splash design on the curling team uniforms totally radiates that traditional Chinese vibe. I hereby declare that domestic Chinese brands have officially won my heart!''. In these dominant narratives, citizens were encouraged to believe that national achievements -- Olympics success, commercial competitiveness, international recognition -- stemmed from regime capabilities and state strength. The underlying message is that this regime makes China great. 

When events challenges regime legitimacy, state narratives pivot to emphasize government responsiveness and problem-solving capacity. Dominant narratives focused on official actions addressing crises. Both the Shanghai Lockdown and Chained Woman incident exhibited this pattern, despite their different threatening mechanisms -- former as policy failure and the latter as the exposure of crucial social problems. In the Shanghai Lockdown, top triplets emphasized collective mobilization and responsive action: “personnel/supplies support Shanghai'' and “Nationwide has (supported) Shanghai'' stressed solidarity and resource deployment, while “Shanghai respond to situation/problem'' and “Shanghai starts nucleic acid/antigen testing'' foregrounded government measures. These narrative elements constructed an image of active, capable governance confronting a difficult situation.

The Chained Woman incident similarly centered official action despite the fundamentally different nature of the threat. Top triplets like “official announce situation,'' “department delivers/sends Yang Mouxia (to hospital for treatment),'' and “Jiangsu/Shaanxi set up investigation team'' documented investigative and remedial steps. The three most influential posts from government and media accounts -- all from the News Center of Chinese Central Television -- emphasized local governments' concrete achievements (“establish investigation team'') and solemn commitments: “uncover the truth, severely punish relevant illegal criminal behavior according to law, hold responsible personnel strictly accountable, and publish the results to the public in a timely manner.'' The core belief these narratives promoted was consistent across threatening events: despite problems, the regime is benevolent, responsive, and capable of solving issues. 

\subsection{Finding 2: Citizen Verification Capacity Constrains Narrative Strategies}
Citizen verification capacity -- whether citizens can verify official narratives against alternative information sources -- constrained how regimes could strategically construct those above-mentioned beliefs. This finding speaks to an ongoing debate in propaganda studies about whether authoritarian messaging operates primarily through top-down manipulation or by affirming and channeling citizens' existing beliefs and emotions \cite{greene2019putin, shirikov2024rethinking,tang2016populist, su2016selective, ding2022performative,greene2022affect}. Our evidence suggests that citizen verification capacity shapes which approach dominates. When citizens can verify state narratives through lived experience or accessible evidence, regimes adopt more bottom-up, responsive strategies that acknowledge citizen-verified realities; when citizens lack verification means, regimes enjoy greater latitude for top-down information manipulation and selective framing.

When citizens could directly verify information through lived experience, regimes faced significant constraints on narrative construction. They could not simply fabricate information or deny visible realities, but instead had to acknowledge citizen-verified facts and demonstrate responsiveness to citizen concerns. The Shanghai Lockdown exemplified these strategic constraints. Shanghai residents directly experienced lockdown conditions, including food shortages, quarantine failures, policy chaos, making official claims immediately verifiable against lived reality. This verification capacity shaped narrative strategies in observable ways. Beyond the emphasis on collective mobilization, responsive action and government measures discussed in Finding 1, government narratives also included factual acknowledgments that citizens could independently verify. The top triplet “Shanghai has new case/death'' simply reported pandemic developments without interpretive framing. This factual reporting represents a strategic concession to verification constraints. 

When citizens lacked independent means to verify official narratives -- due to geographic distance, information control, technical or language barrier -- regimes can exercise far greater control over narrative construction and framing. They could selectively present information, introduce unverified claims, and construct causal interpretations. The Russia-Ukraine War clearly demonstrated this pattern. Chinese citizens had no direct experience of the conflict, the Great Firewall blocked access to international news sources and alternative perspectives, and geographic distance and language barrier prevented independent assessment. Under these conditions, the regime has considerably more leeway to construct belief. Both government/media accounts and celebrity accounts constructed a consistent narrative dominated by three key actors: Russia, Ukraine and the US. The top triplets establishes a carefully crafted causal sequence framing the war's origins and ongoing dynamics. Three strategic patterns reveal how low citizen verification capacity enabled narrative manipulation. 

First, causal attribution chains linked the key actors in ways that justified Russian actions: “Ukraine joined NATO,'' “Russia invaded Ukraine,'' “the US imposed sanction on Russia.'' These dominant relationships construct a narrative sequence that contextualizes Russian military action as a response to geopolitical provocation arising from Ukraine's NATO aspirations, with subsequent US sanctions portrayed as further evidence of Western aggression. This causal framing, possible because citizens could not independently verify the complex diplomatic history, deflected responsibility from Russia to Western actors.

Second, unverified action attributions introduced contested claims as fact: “the US conducts biological experiment.'' The biological experiments allegation is unverified and contested internationally, yet appeared prominently in top triplets without qualification or evidence. This pattern reveals how low citizen verification capacity enables regimes to introduce unsubstantiated allegations that cast geopolitical rivals in threatening terms -- a strategy unavailable when citizens can fact-check claims.

Third, moral framing redirected culpability through interpretive claims: “the West abandon Ukraine''; “the US implements genocide.'' These dominant triplets construct a counter-narrative to Western condemnation of Russia's invasion, portraying Western actors, particularly the United States, as morally culpable. Without access to alternative sources, citizens encountered this moral framing as the dominant interpretation.

These strategic patterns manifested in widely circulated posts. China Central Television (CCTV) News reported on Russia's justification for its military operation -- Ukraine joining the NATO. Government and media accounts actively shared CCTV's posts that explicitly attributed the war to US actions. Posts on US sanctions against Russia generated significant engagement on Weibo, including a popular hashtag “Western Sanctions Target Russian Civilians'', which framed Western economic measures as deliberately harming ordinary citizens rather than Russia government officials. This narrative construction -- causal chains placing blame on the West, unverified allegations against the US, moral condemnation of Western actors -- was possible precisely because citizens lacked means to verify these claims against alternative information.

\subsection{Finding 3: Beyond Simple Patterns, The Complexities in Authoritarian Storytelling}

Findings 1 and 2 established that legitimacy implications shape belief content while verification capacity constrains strategies of state narratives. However, real world events rarely present regimes with clear-cut conditions and follow fixed patterns, they often exhibit mixed and shifting properties and occupy ambiguous positions. We discuss these complexities and their implications in authoritarian storytelling in this section. 

First, dimensional combinations shape storytelling challenges. We can conceptualize these challenges along a continuum from maximum constraint to maximum freedom. At one extreme, events likely to challenge regime legitimacy with high citizen verification capacity create the most constrained environment for the construction of state narratives. The regime needed to restore legitimacy (requiring benevolent leadership narratives) while facing citizens who could verify claims through lived experience. Shanghai Lockdown represents such a combination where we observed a constrained responsiveness. State narratives acknowledged undeniable facts, emphasized mobilization and government action, attempting to cultivate belief in a responsive and benevolent state. At the opposite extreme, events likely to boost regime legitimacy with low citizen verification capacity provides maximum freedom for propaganda. The Beijing Winter Olympics exemplifies this combination. The state was able to amplify nationalist achievement without the constraints of citizen verification capacity. We observed an unconstrained nationalism that left little room for criticism. Through state storytelling, success was celebrated by linking domestic commercial brands to national pride and by connecting Olympic accomplishments and global recognition to regime capability. At the core of this narrative lies the belief in a “nationalist China” -- a vision of greatness attainable only under the current government’s leadership and capacity to mobilize the nation.

Between these extremes, other combinations create challenging of varying levels. Generally, events that are likely to threaten regime legitimacy tighten the construction of state narratives more than those that are likely to bolster it, while events with high citizen verification capacity impose greater constraints than those with low capacity. For example, the state faced considerable more constrain in crafting its narratives in the Chained Woman incident than in Russian Ukraine War, as the Chained Woman incident combined both a relatively higher citizen verification capacity and greater reputation damage to regime legitimacy. 

Second, dimensions are not fixed context, but strategic variables regimes can manipulate. Citizen verification capacity proves particularly malleable: regimes can reduce citizen access to information through censorship and attention diversion, well documented in the literature of authoritarian propaganda \cite{king2013censorship, king2017chinese,munger2019elites}, thereby expanding their available storytelling strategies. The Chained Woman incident illustrates this dynamic. Initially, the event exhibited high citizen verification capacity. Influencers exposed the case through viral videos that drew widespread public attention, prompting many citizens and independent journalists to travel to Fengxian to document and report on the Chained Woman incident. Under these high-verification conditions, the regime faced storytelling constraints similar to the Shanghai Lockdown: it had to acknowledge the incident, announce investigations, and demonstrate responsive action (Finding 1).

However, as the issue revealed deep-seated social problems -- human trafficking and systematic failures that threatened regime legitimacy -- the regime imposed censorship and attention diversion online and local restrictions, prohibiting further interviewing and reporting by anyone other than state media. The censorship and local restrictions effectively reduced citizen verification capacity over time. The effect of online censorship is evident in the number of posts: only 3,316 posts were identified for this event, compared with at least 149,526 posts for each of the other three events examined, reflecting both the suppression of online discussion and the success of attention diversion efforts.

This dynamic reveals that citizen verification capacity is not merely a constraint regimes have to adapt to, but a variable that can be strategically modified. Yet, the manipulation of citizen verification capacity has obvious limits. In the Chained Woman incident, which was first exposed by influencers online, the regime could not retroactively erase the initial viral videos that millions had already viewed, nor could it completely eliminate public awareness of the case. Subsequent censorship and attention diversion reduced citizens' ongoing ability to verify official information and state narratives, but because the event was already widely known, the state was compelled to engage in storytelling rather than relying solely on suppression. Moreover, extensive censorship ultimately undermine media credibility and, more fundamentally, political trust \cite{GEHLBACH2014163,StockmannLuoforthcoming}.

Third, the legitimacy implications of events are not always crystal clear; in some cases, they are mixed or ambiguous, particularly in cases involving geopolitical conflicts, which complicates the belief that state narratives seek to cultivate. The Russia-Ukraine War exemplifies this challenge. China's strategic partnership with Russia and emphasis on sovereignty made condemning the invasion diplomatically costly, yet mounting international criticism risked political isolation. Our analysis revealed that in such contexts, the belief promoted by state narratives pivots on identity-based nationalism, which defines in-group and out-groups -- an “us-versus-others'' logic. In the case of Russia Ukraine War, causal attribution chains that justified Russian invasion as a defense of sovereignty and shifted moral culpability from Russia to the US aligned with the identity-based nationalism that the Chinese state seeks to instill among citizens. 

This represents a different dimension of nationalist belief construction. The Olympics emphasized patriotic nationalism -- pride in national achievement and regime capability. The Russia-Ukraine War, by contrast, emphasized victimization nationalism -- resentment toward other nations stemming from perceived historical victimization and injustices \cite{WoodsDickson2017}. The Chinese state has long cultivated victimization narratives portraying China as historically humiliated by Western powers and Japan since the Opium Wars \cite{WoodsDickson2017, xu2023power}. The Russia-Ukraine narratives tapped into this victimization frame, constructing an “Us-versus-West'' identity that positioned China alongside Russia against Western aggressors. Similar patterns appear in other authoritarian contexts. Serbian regimes have constructed narratives of the Yugoslav secession wars that unite populations against common enemies \cite{subotic2013remembrance}, while Russian state media coverage of Ukraine similarly emphasizes “Us-versus-West'' identity narratives \cite{Bradshawetal2024}. Research demonstrates that such collective identity narratives effectively increase suspicion towards foreign governments and reinforce popular support for ruling regimes \cite{xu2023power}.

\section{Conclusion}

In this article, we examine how authoritarian regimes construct state narratives about politically consequential events on social media. We build on the narrative policy framework and borrow two dimensions from existing literature on authoritarian propaganda as the analytical lenses for our exploratory analysis: legitimacy implications, and citizen verifiability capacity. Our findings reveal that legitimacy implications shape the beliefs that state narratives seek to cultivate, while citizen verification capacity constraints the range of narrative strategies available to the state. When events enhance regime legitimacy, state narratives link national pride and achievement with the state, reinforcing a belief in a nationalist and great country attainable only under the current state's leadership. When events threaten legitimacy, narratives pivot toward themes of responsiveness and problem-solving, emphasizing the benevolent and responsive leadership. Citizen verification capacity balances the top-down manipulation and bottom-up responsiveness in state narratives. When citizens can verify state narratives through lived experience or accessible evidence, regimes adopt more bottom-up, responsive strategies that acknowledge citizen-verified realities and emphasize mobilization and government actions; when citizens lack verification means, regimes enjoy greater latitude for top-down information manipulation and selective framing. Our findings reveal propaganda as a complex and adaptive process of narrative construction.

We acknowledge two limitations of our approach. One limitation concerns the interpretation of extract triplets. When we identify sample sentences for dominant triplets, it is possible that some have been hijacked or repurposed by users, particularly celebrities, for criticism, satire, or unrelated topics, for instance, when popular hashtags or official phrases are reused in divergent contexts. While such instances introduce semantic noise, they do not undermine our central claim: the prominent of particular triplets still reflect the salience and visibility of their core meaning as state narratives. In other words, even with noise, these dominant triplets reveal the narrative strategies and promoted beliefs in authoritarian storytelling. Second, the automated extraction of triplets from Chinese texts inevitably introduces some noise due to linguistic ambiguity and parsing errors. We mitigated these risks through manual validation of key triplets and cross-checks with representative posts, but residual error remains possible. Despite these limitations, the method provides a scalable and systematic approach to identifying dominant narrative patterns and enables a qualitative zoom-in on how these patterns structure authoritarian propaganda.

These findings contribute to broader debates on authoritarian resilience, underscoring that information control is not merely repressive or top-down but also adaptive and strategic. Adaptation in authoritarian storytelling about politically consequential events emerges as a key survival strategy, enabling regimes to recalibrate state narratives in response to shifting legitimacy implications and varying levels of citizen verification capacity. This dynamic process helps explain why some authoritarian regimes continue to enjoy high levels of political support \cite{chen2016sources, chapman2021shoring, tang2016populist}, as propaganda not only asserts top-down control but also affirms citizens' existing beliefs and emotions — a pattern consistent with existing research on authoritarian propaganda. Future studies could apply this framework cross-nationally to examine the art of authoritarian storytelling in different contexts. While it is beyond the scope of this article to assess citizen reception, our findings suggest that when citizen verification capacity is high, propaganda becomes more responsive. Future work could therefore extend this framework by integrating citizen reception and feedback as an additional dimension in understanding how propaganda operates as an evolving dialogue between state and society.

\clearpage
\begin{landscape}
\begin{center}
\setlength\LTleft{0pt}
\setlength\LTright{0pt}

\begin{xltabular}{\textwidth}%
  {>{\raggedright\arraybackslash}p{1.8cm}
   >{\raggedright\arraybackslash}p{1.8cm}
   >{\raggedright\arraybackslash}p{1.8cm}
   >{\centering\arraybackslash}p{1cm}
   >{\raggedright\arraybackslash}p{14.6cm}}
\caption{Top 5 Triplets for Different Events} \\
\toprule
ARG0 & B-V & ARG1 & counts & Sample sentence(s) \\
\midrule\midrule
\endfirsthead

\toprule
ARG0 & B-V & ARG1 & counts & Sample sentence(s) \\
\midrule
\endhead

\bottomrule
\endfoot

\multicolumn{5}{l}{Russia Ukraine War: Government/media accounts} \\
\midrule
USA & implements sanction on & Russia & 349 &
\begin{minipage}[t]{\linewidth}
...the United States and its allies have imposed 8,068 sanctions on Russia over eight years... \\[4pt]
...U.S. Secretary of State Blinken announced visa sanctions against Russian officials.\\[4pt]
The United States will impose sanctions on Russia’s financial sector.
\end{minipage} \\
\addlinespace

Russia & invades & Ukraine & 278 &
\begin{minipage}[t]{\linewidth}
Biden said that as Russia continues its invasion of Ukraine, this move will deliver “another crushing blow” to Russia. \\[4pt]
Tens of thousands of Serbians waved Russian flags and carried photos of President Putin as they marched in Belgrade toward the Russian embassy—a rare display of public support for Moscow following its invasion of Ukraine. \\[4pt]
U.S. President Biden stated on the 17th (local time) that he believes Russia will invade Ukraine in the coming days.
\end{minipage} \\
\addlinespace

Russia & takes & action/ operation & 261 &
\begin{minipage}[t]{\linewidth}
..the German government’s official stance on Russia’s special military operation in Ukraine...\\[4pt]
Russia takes action in response to Western economic sanctions.\\[4pt]
Since Russia launched its military operation against Ukraine...
\end{minipage} \\
\addlinespace

Ukraine & joined/ belonged & NATO & 206 &
\begin{minipage}[t]{\linewidth}
Ukrainian President Zelensky delivered a video speech, stating that the West has completely abandoned Ukraine: “I asked 27 European leaders whether Ukraine would join NATO?”\\[4pt]
\#Kissinger’sWarning8YearsAgoOnTheRussiaUkraineIssue\# \#HowShouldTheRussiaUkraineIssueEnd\# In 2014, when conflict also broke out between Russia and Ukraine, former U.S. Secretary of State Henry Kissinger wrote: if Ukraine is to survive and develop, ... it should not join NATO.\\[4pt]
Former U.S. Congresswoman Tulsi Gabbard stated that Biden could end this crisis and prevent war simply by promising not to admit Ukraine into NATO—but they refuse to do so.
\end{minipage} \\
\addlinespace

West & abandon & Ukraine & 175 &
\begin{minipage}[t]{\linewidth}
\#UkrainianPresidentSaysTheWestHasCompletelyAbandonedUkraine\# \\[4pt]
On the 25th (local time), Ukrainian President Zelensky released a video speech, stating that the West has completely abandoned Ukraine: “I asked 27 European leaders whether Ukraine would join NATO? But everyone was afraid, no one answered me.”
\end{minipage} \\
\midrule

\multicolumn{5}{l}{Russia Ukraine War: Celebrity accounts} \\
\midrule
Russia & invades & Ukraine & 249 &
\begin{minipage}[t]{\linewidth}
..Russia’s invasion of Ukraine is not because of NATO’s eastward expansion... \\[4pt]
...China has not yet condemned Russia’s “invasion” of Ukraine...\\[4pt]
...Russia’s violent invasion of Ukraine has been going on for over a month...
\end{minipage} \\
\addlinespace

USA & impose sanction on & Russia & 179 &
\begin{minipage}[t]{\linewidth}
\#IntelAppleGoogleJoinUSSanctionsAgainstRussia\#\\[4pt]
The United States continues to escalate unilateral sanctions against Russia and coerces the world to take sides.\\[4pt]
The U.S. doesn’t care about the lives of the Ukrainian people; its goal is simply to find excuses to sanction and collapse the Russian economy.
\end{minipage} \\
\addlinespace

USA & implements & genocide & 151 &
\begin{minipage}[t]{\linewidth}
Full text! The Ministry of Foreign Affairs released the “Historical Facts and Real Evidence of the U.S. Implements Genocide Against Native Americans”.\\[4pt]
Demons walk among humans—that’s America. {[}Full text! The Ministry of Foreign Affairs released “Historical Facts and Real Evidence of the U.S. Implements Genocide Against Native Americans.”{]}
\end{minipage} \\
\addlinespace

Ukraine & joined/ belonged & NATO & 97 &
\begin{minipage}[t]{\linewidth}
…It’s a fact that Ukraine wanted to join NATO...\\[4pt]
After Zelensky dragged Ukraine into war, he finally admitted that Ukraine would not join NATO.\#ZelenskySaysUkraineNowUnderstandsItCannotJoinNATO\#\\[4pt]
Putin says Ukraine joining NATO is completely unacceptable.
\end{minipage} \\
\addlinespace

USA & conducts/ has & biological experiment & 85 &
\begin{minipage}[t]{\linewidth}
Latest news—the Russian Ministry of Defense says the U.S. established and funded biological laboratories in Ukraine where experiments were conducted on bat coronavirus samples.\\[4pt]
\#BehindTheUSBiologicalLaboratories\# Behind America’s biological laboratories lies inhumanity, lawlessness, moral decay, and utter depravity—its crimes are beyond measure.
\end{minipage} \\
\midrule

\multicolumn{5}{l}{Beijing Winter Olympics: Government/media accounts} \\
\midrule
Beijing & has new/ reports & patients/ infections/ cases & 236 &
\begin{minipage}[t]{\linewidth}
\#BeijingReports7NewInfections\#\\[4pt]
\#BeijingNewCasesIncludeBusinessTravelReturneesAndTheirFamilies\#\\[4pt]
\#BeijingReports9ConfirmedCasesAnd3Asymptomatic\#
\end{minipage} \\
\addlinespace

China & participates & Winter Olympics/Beijing Winter Paralympics Competition & 157 &
\begin{minipage}[t]{\linewidth}
\#OnThisDay42YearsAgoChinaFirstParticipatedInTheWinterOlympics\# From winning its first medal to its first gold medal, and now, 42 years later, Beijing has become the “Dual Olympic City'', China's ice and snow sports have never stopped moving forward.\\[4pt]
From China’s first participation in Winter Olympics in 1980 to hosting its first Winter Olympics in 2022, these years have seen not only “zero breakthroughs” in gold medal counts for various events, but also a growing national passion for the Winter Games.
\end{minipage} \\
\addlinespace

World & expects/ awaits & China & 108 &
\begin{minipage}[t]{\linewidth}
In one month, the Beijing Winter Olympics will officially open \#OpenTheBeijingWinterOlympicsWithChineseStyleTrailer\# Gliding by the Great Wall, soaring in factories, blooming above the Bird's Nest... the world awaits China — and China is ready.\\[4pt]
Editorial from Southern Daily: Building the Ice-and-Snow Dream Together, Toward the Future — Written at the Opening of the Beijing Winter Olympics: “The world awaits China, and China is ready.”\\[4pt]
The Beijing Winter Olympics and Paralympics Summary and Commendation Ceremony was held on the morning of April 8 in the Great Hall of the People. “Two Olympics, Equally Splendid” — China did not disappoint the world's expectation and delivered a satisfactory answer sheet.
\end{minipage} \\
\addlinespace

China & is & prepared/ready & 99 &
\begin{minipage}[t]{\linewidth}
...curtain of the Beijing Winter Olympics is about to open — China is ready. \\[4pt]
Xi Jinping: “The Beijing Winter Olympics will open tomorrow evening — China is ready.”
\end{minipage} \\
\addlinespace

State Council & wish the best/ congratulates & Beijing & 89 &
\begin{minipage}[t]{\linewidth}
Congratulatory Message from the CPC Central Committee and the State Council to the Chinese Sports Delegation of the 24th Beijing Winter Olympics\\[4pt]
Congratulatory Message from the CPC Central Committee and the State Council to the Chinese Sports Delegation of the 24th Beijing Winter Paralympics
\end{minipage} \\
\midrule

\multicolumn{5}{l}{Beijing Winter Olympics: Celebrity accounts} \\
\midrule
China & participates & Winter Olympics/ Beijing Winter Paralympics Competition & 157 &
\begin{minipage}[t]{\linewidth}
The Chinese delegation at the Beijing Winter Paralympics participated in all major events and currently ranks first on both the gold medal and total medal tables — a historic achievement.\\[4pt]
Repost Weibo: \#OnThisDay42YearsAgoChinaFirstParticipatedInTheWinterOlympics\#... \#Beijing2022WinterOlympics\# is underway — let’s move toward the future together!
\end{minipage} \\
\addlinespace

China & has & ANTA Sports & 79 &
\begin{minipage}[t]{\linewidth}
Whether it’s @EileenGu’s 1440° flip on the slopes or @UNIQ-WangYibo’s Ollie trick on the street — on skis or skateboards, just go for it! The Beijing Winter Olympics awaits your challenge! Gu Ailing, Wang Yibo \#DareToSlide\# \#ChinaLovesSportsChinaHasANTAsSports\# \#WinterOlympics\# ANTA’s Weibo video\\[4pt]
\#ChinaLovesSportsChinaHasANTAsSports\# Watch the Beijing Winter Olympics, feel the speed on ice — go, Team China! Can’t wait for that championship moment!\\[4pt]
I’m taking part in ANTA’s Winter Olympics brand challenge — complete the missions to unlock exclusive Winter Olympics merchandise! Join me! \#BrandMoment\# \#WinterOlympics\# ANTA Come join in! “Loves Sports, China Has ANTA Sports.”
\end{minipage} \\
\addlinespace

World & expects/ awaits & China & 51 &
\begin{minipage}[t]{\linewidth}
Countdown 10 days to the Winter Olympics — the world awaits China, and China is ready! Let’s imagine the future through Zhang Jie’s song — see you in Beijing! \\[4pt]
On February 4, the world’s eyes turn to the Beijing 2022 Winter Olympics as they open. Xinhua News Agency launches the original Winter Olympics-themed MV “A Promise to the World,” sung by Li Yuchun. “The world awaits China, and China is ready.” A winter celebration, a spring invitation.
\end{minipage} \\
\addlinespace

Beijing & has new/ reports & patients & 50 &
\begin{minipage}[t]{\linewidth}
\#BeijingChaoyangDistrictReports2NewPositiveCasesToday\# I don’t even care if we have classes anymore, I’m exhausted — as a 2019 student, I’ve basically turned my undergrad into a vocational diploma. \\[4pt]
\#BeijingReported48NewLocalCasesYesterday\# Beijing is getting ridiculous. It’s a 10-minute walk from my apartment to campus, and I saw three nucleic acid testing sites along the way — not even counting the one inside school.\\[4pt]
\#BeijingReported36NewLocalCasesYesterday\# Stay strong, Beijing — it’ll get better soon.
\end{minipage} \\
\addlinespace

China & is & prepared/ ready & 49 &
\begin{minipage}[t]{\linewidth}
@//ZhaoLijianOfficialWeibo: Together Toward The Future! \#LatestWinterOlympicsPromoReleased\# \#BeijingWinterOlympicsOpenTomorrow\#! ...The world awaits China — and China is ready. Welcome, friends from all over the world — let’s move toward the future together!
\end{minipage} \\
\midrule

\multicolumn{5}{l}{Shanghai Lockdown: Government/media accounts} \\
\midrule
Personnel/ Supplies & support & Shanghai & 1595 &
\begin{minipage}[t]{\linewidth}
Shanghai's battle against COVID-19 touches everyone's heart! The medical team from Shanxi has arrived!\#Over38000MedicalStaffFrom15ProvincesRushedToSupportShanghai\# \#ThumbsUpForShanxi\#\\[4pt]
\#InnerMongoliaAirlifts400TonsOfBeefMuttonAndPorkToSupportShanghai\#: TwoFlightsPerDay, CompletedIn5Days\\[4pt]
\#Over2000TestingPersonnelFromJiangsuAndZhejiangArriveInShanghaiToAssist\# Nucleic acid testing teams from Jiangsu and Zhejiang provinces are currently assisting Shanghai Pudong in a new round of mass testing.
\end{minipage} \\
\addlinespace

Shanghai & has new/ reports & case/ death & 522 &
\begin{minipage}[t]{\linewidth}
\#ShanghaiReports994NewLocalCasesAnd22348Asymptomatic\#  First Batch Of “ThreeZones” Division List Released\\[4pt]
\#ShanghaiReported358NewLocalCasesYesterday\# \#4144LocalAsymptomaticInfections\#\\[4pt]
\#ShanghaiReports914LocalCasesAnd25173Asymptomatic\# LiangWannian: We Must Not Take The Shanghai Epidemic Lightly\#CentralizedQuarantineStillNecessaryUnderCurrentConditions\#
\end{minipage} \\
\addlinespace

Shanghai & takes & nucleic acid/antigen testing & 180 &
\begin{minipage}[t]{\linewidth}
\#ShanghaiConductsNewRoundOfCitywideNucleicAcidTesting\# On April 4, Shanghai carried out citywide nucleic acid testing.\\[4pt]
\#ShanghaiLockdownAreasWillUndergoThreeConsecutiveDaysOfTesting\#\\[4pt]
Since April 15, Shanghai has completed over 37 million nucleic acid tests. Starting today, three consecutive days of testing will take place in lockdown areas, and on the 20th, testing will also begin in controlled areas.
\end{minipage} \\
\addlinespace

Epidemic & is in & development & 157 &
The COVID-19 situations in Shanghai and Jilin are still developing; receiving three vaccine doses can effectively reduce mortality after Omicron infection. \\
\addlinespace

Nation-wide & has (supported) & Shanghai & 81 &
\begin{minipage}[t]{\linewidth}
\#ShanghaiBuildsLargestMakeshiftHospitalIn184Hours\# \#PfizerCOVIDOralDrugNowUsedIn8Provinces\#  \#Over10000MedicalStaffNationwideSupportShanghai\#\\[4pt]
\#Over10000MedicalStaffNationwideSupportShanghai\# \#HainanMedicalTeamSupportsShanghai\# \#HainanAssistsShanghai\#
\end{minipage} \\
\midrule

\multicolumn{5}{l}{Shanghai Lockdown: Celebrity accounts} \\
\midrule
Personnel/ Supplies & support & Shanghai & 1701 &
\begin{minipage}[t]{\linewidth}
\#ShanghaiEpidemic\# On April 8, the Joint Logistics Support Force’s 990th and 989th hospitals arrived in Shanghai. \#People’sArmySupportsShanghai\#\\[4pt]
Though not widely reported, please remember — there are also 1,500 medical staff from Gansu supporting Shanghai. \\[4pt]
\#DriverTravelsOver5000KmToDeliver10TonsOfXinjiangMuttonToShanghai\#
\end{minipage} \\
\addlinespace

Shanghai & has new/ reports & case/ death & 516 &
\begin{minipage}[t]{\linewidth}
\#ShanghaiReports3084NewLocalCasesAnd17332Asymptomatic\# \#7NewLocalDeaths\#\\[4pt]
\#ShanghaiReports62NewLocalAsymptomaticCases\# Last night, the nearby middle school ran nucleic acid testing overnight.\\[4pt]
Some buildings in my neighborhood are locked down — finally feeling the outbreak right next door {[}cry{]} Please let this end soon, I’m begging…
\end{minipage} \\
\addlinespace

National & has (supported/ aided) & Shanghai & 222 &
\begin{minipage}[t]{\linewidth}
\#Over10000MedicalStaffNationwideSupportShanghai\# All flights have stopped; those arriving now are all medical personnel coming to help Shanghai.\\[4pt]
“United in purpose, honoring the mission! Determined to complete the task!” The Shandong medical team aiding Shanghai — heroes!\\[4pt]
According to Xinhua News, medical teams from Tianjin, Hubei, Hainan, Jiangxi, Shandong, Henan, Anhui, and other regions \#Over10000MedicalStaffNationwideSupportShanghai\#
\end{minipage} \\
\addlinespace

Shanghai & response to & situation/ problem & 140 &
\begin{minipage}[t]{\linewidth}
\#ShanghaiWomen’sFederationRespondsToIssueOfInfantsBeingIsolatedAlone\#\\[4pt]
\#ShanghaiExplainsWhyPositiveCasesStillAppearInLockedDownCommunities\# Why are positive cases still emerging even after long lockdowns? Shanghai gave three reasons in response.\\[4pt]
\#ShanghaiRespondsToBuyingGroceriesInBeicaiPudong\#
\end{minipage} \\
\addlinespace

Shanghai & starts/ conducts & nucleic acid/ antigen testing & 91 &
\begin{minipage}[t]{\linewidth}
\#ShanghaiTesting\# Shanghai, as the hardest-hit area this time, finally starts citywide nucleic acid testing. \\[4pt]
\#ShanghaiConductsNewRoundOfCitywideTesting\# and once again, the whole nation \#SupportsShanghai\#!\\[4pt]
\#ShanghaiConductsNewRoundOfCitywideTesting\# Why wasn’t this done earlier? Dragged the entire city and medical staff for a whole month, wasting manpower and resources.
\end{minipage} \\
\midrule

\multicolumn{5}{l}{The Chained Woman event: Government/media accounts} \\
\midrule
Fengxian & give birth to & woman with eight kids & 549 &
\begin{minipage}[t]{\linewidth}
…netizens reported information related to the “woman who gave birth to eight children.”\\[4pt]
A video sparked heated discussion online; netizens said a woman with mental illness had given birth to 8 children. It was filmed in Feng Xian, Xuzhou, Jiangsu Province. \\[4pt]
\#OfficialBriefingOnXuzhouFengXianWomanWhoGaveBirthToEightChildren\#
\end{minipage} \\
\addlinespace

Official/ Xuzhou/ Fengxian/ Investigation Team/ Procuratorate & announce/ reports & situation & 27 &
\begin{minipage}[t]{\linewidth}
Official response on the Jiangsu mother of eight being chained in a dilapidated hut.\\[4pt]
The Jiangsu Provincial Party Committee \& Provincial Government investigation team releases a notice on the investigation and handling of the “Feng Xian woman who gave birth to eight children” incident... \\[4pt]
\#XuzhouReportsHandlingInFengXianEight-ChildCase\#: The “husband” and 2 others have been placed under criminal compulsory measures
\end{minipage} \\
\addlinespace

Yang Mouxia & is & Xiao Hua Mei & 25 &
\begin{minipage}[t]{\linewidth}
Yang Mouxia’s original name is Xiaohuamei.\\[4pt]
\#FengXianEight-ChildWomanConfirmedAsXiaohuamei\#
\end{minipage} \\
\addlinespace

Department & deliver/ send & Yang Mouxia & 22 &
\begin{minipage}[t]{\linewidth}
On Jan 28, 2022,  the health departments of Xuzhou and Feng Xian sent Yang Mouxia to hospital for examination and treatment \\[4pt]
\#XuzhouCitywideInspectionOnViolationsOfWomenAndChildren’sRights\# On Jan 28, 2022, Xuzhou and Feng Xian health departments sent Yang to hospital; staff and her eldest son accompanied her. Treatment is ongoing.
\end{minipage} \\
\addlinespace

Result & is & Pu Mouma and Yang Mouxia & 20 &
On Feb 9, 2022, the Ministry of Public Security Forensic Identification Center conducted DNA testing comparing blood samples from Yang Mouxia and Guang Mouying (Xiaohuamei’s half-sister; original name Hua Mouying) with biological materials extracted from the belongings of Pu Mouma (Xiaohuamei’s mother, deceased in 2018). \\
\midrule

\multicolumn{5}{l}{The Chained Woman event: Celebrity accounts} \\
\midrule
Fengxian & give birth to & woman with eight kids & 769 &
\begin{minipage}[t]{\linewidth}
When it was reposted 104,000 times, it was also gone. Are you happy this spring? I can’t be happy. \#XuzhouPublishesProgressOnFengXianEight-ChildCase\# \\[4pt]
Strongly demand a thorough investigation! Are there other illegal/criminal acts?   The people want the investigation team to dig to the bottom! \\[4pt]
This is the closest these women have come to being rescued!!! \#FengXianEight-ChildCaseInvestigationTeamFormed\# \#OfficialBriefingOnXuzhouFengXianEight-ChildCase\# \#XuzhouPublishesProgressOnFengXianEight-ChildCase\#
\end{minipage} \\
\addlinespace

Official/ Xuzhou/ Fengxian/ Investigation Team & announces/ reports & situation & 651 &
\begin{minipage}[t]{\linewidth}
\#XuzhouReleasesProgressOnFengXianEight-ChildInvestigation\# “The scariest part of ‘abduction and trafficking’ is not the ‘abduction’ but the ‘trafficking.’ Abduction might be a low-probability sudden event, but ‘trafficking’ implies buyers, sellers—even an entire industry chain… That’s what’s most terrifying.”
\end{minipage} \\
\addlinespace

Jiangsu & set up/ form & investigation team & 60 &
\begin{minipage}[t]{\linewidth}
Finally! \#JiangsuProvinceFormsInvestigationTeamForFengXianEight-ChildCase\#\\[4pt]
Hope a central (national-level) investigation team is formed...\\[4pt]
Don't let the public down! \#JiangsuProvinceFormsInvestigationTeamForFengXianEight-ChildCase\#
\end{minipage} \\
\addlinespace

Shaanxi & set up/ form & investigation team & 30 &
\begin{minipage}[t]{\linewidth}
\#ShaanxiFormsInvestigationTeamFor“IronCageWoman”Incident\# \\[4pt]
Even if I’m shadow-banned, muted, or have my account nuked, I will still care. \#ShaanxiFormsInvestigationTeamForIronCageWoman\#\\[4pt]
Any progress? \#FengXianEight-ChildCaseInvestigationTeamFormed\# How is Xiaohuamei’s treatment going and what will happen next?
\end{minipage} \\
\addlinespace

Lawyer & comments on & Xuzhou & 19 &
As a criminal defense lawyer used to scrutinizing dozens or hundreds of volumes of files… the Feb 7 report’s reasoning is wrong, it dodges key issues, and its conclusions cannot stand... \\
\end{xltabular}

\end{center}
\end{landscape}

\clearpage

\section*{Acknowledgments}
This research was supported by grants from the British Academy (SRG21/210233). We are grateful to Koen P.R. Bartels, Peter Gries, Ning Leng, René Lindstädt, Lizhi Liu, Bingchun Meng, Dion Stevers, and Daniela Stockmann, as well as the three anonymous reviewers and the editors, for their constructive feedback and insightful suggestions, which greatly improved this paper. We also thank Anfan Chen for making the data collection possible. We further appreciate the feedback from participants at the American Political Science Association Annual Meeting (August 2023 \& September 2024), the Department of Social Statistics Research Seminar at University of Manchester (Nov 2023), the 6th Annual COMPTEXT Conference (May 2024), the International Society for Computational Social Science (July 2024), the Chinese Politics and Economy Research Seminar Series at Georgetown University (December 2024), the China and the World Seminar Series at LSE (March 2025), the Department of Public Administration and Policy Writing Retreat at the University of Birmingham (Spring 2025),  and the Political Studies Association Annual Conference (April 2025) for their constructive comments on earlier versions of this paper.

\bibliographystyle{unsrt}  
\bibliography{references}

\clearpage

\begin{appendix}
\setcounter{figure}{0}
\renewcommand\thefigure{A\arabic{figure}} 
\setcounter{table}{0}
\renewcommand\thetable{A\arabic{table}}

\section{Data Collection and Processes}

The dataset was collected from the active user pool constructed by \cite{hu-etal-2020-weibo}, which includes approximately 20 million active users on Weibo (accounting for 8 percent of all Weibo users). They defined active Weibo users as meeting the following two criteria: (1) The number of followers, fans, and posts are all greater than 50; (2) The most recent post was made within the last 30 days. We then collected all the posts made by a random sample of 300,000 users during the five months between January 1st and May 5th, 2022. Among these users, we filtered for those verified by the Weibo platform, resulting in 38,019 verified users across 7 categories. The following table shows the distribution of the verified users and their posts.

\begin{table}[!ht]
\centering
\caption{Description of verified user and posts data}
\begin{tabular}{p{1.7cm}|p{1.2cm}|p{1.2cm}|p{1.2cm}|p{0.9cm}}
\hline \hline 
 & N of users & N of posts & User ratio & Post ratio \\\hline
Celebrities & 31112 & 2113421 & 0.818 & 0.637 \\ \hline
Government & 2316 & 589110 & 0.061 & 0.178 \\\hline 
Enterprise & 3240 & 203344 & 0.085 & 0.061 \\\hline 
Media & 354 & 257046 & 0.009 & 0.077 \\\hline 
Campus & 505 & 63548 & 0.013 & 0.019 \\\hline 
Website & 1 & 17 & -- & -- \\\hline 
Groups & 491 & 88545 & 0.013 & 0.026 \\\hline 
 Total & 38019 & 3315031 & ~1 & ~1 \\\hline \hline 
\end{tabular}
\end{table}

\subsection{System pipeline}

In Figure A1, we present the flowchart of the data processing. Using the input of all the filtered verified users and their posts, we first clean the data by removing duplicates and perform initial text preprocessing, such as removing special symbols, numbers, and stopwords. Next, we apply NMF topic modeling and select the relevant event-revealing topics. Based on these topics, we build a plain-text corpus focused on the identified issues. With this corpus, we conduct further text preprocessing using the RELATIO package, and parse the segmented sentences with HanLP. Once the semantic roles are labeled, we extract the key subject-verb-object triplets (SVOs) for each sentence and compile the event triplets into a triplet dictionary. Domestic events that were not initiated by the state were identified using keyword searches. We used two key phrases: “Fengxian'' and “the chained woman.'' 

Before visualizing the output, we applied supervised dimensional reduction. We first performed an in-depth reading of the SVOs identified from a 10 percent sample of each event. Based on this analysis, we created a dictionary for synonym replacement and identified patterns in triplets that could be clustered into similar narratives. For instance, `Shanghai-add-new patients/deaths', `Shanghai-had-new affected', and `Shanghai-increased-patients/deaths' are grouped together as `Shanghai-has-new patients/deaths'. In the case of `US-sanction-Russia', the grouped SVOs include `USA-sanction-Russia', `Developing countries-sanction-Russia', `Economic sanction-on-Russia', `US and allies-sanction-Russia', `US-threats-sanction', `US-wield-sanction tool' etc. We then apply the dictionary and patterns to the full corpus.

\begin{figure}
\centering
    \begin{tikzpicture}
    [node distance=10mm, >=latex',
 block/.style = {draw, rectangle, minimum height=8mm, minimum width=20mm,align=center},
rblock/.style = {draw, rectangle, rounded corners=0.5em},
oblock/.style = {ellipse, draw, inner xsep=#1},
tblock/.style = {draw, trapezium, minimum height=5mm, 
                 trapezium left angle=75, trapezium right angle=105, align=center},]
    \node [oblock]    (data)     {Filtered verified user posts};
    \node [rblock, below=of data]  (preprocess1)   {Data cleaning and text preprocessing};
    \node [rblock, below=of preprocess1]  (topics)   {NMF topic modelling};
    \node [block, below=of topics]   (subset)   {Subset based on \\ different issues};
    \node [block, right=of subset]   (corpus)  {Plain-text Corpus};
    \node [block, below=of corpus]    (preprocess2) {Further text preprocessing};
    \node [block, left=of preprocess2]     (parser)  {Parser (hanlp)};
    \node [block, below=of parser]    (extract)   {Extract SVOs};
    \node [block, right=of extract]     (triplets) {Construct event triples \\ and store in \\ triple dictionary};
    \node [block, below=of triplets]  (dimensionreduction)     {Dimension Reduction};
    \node [oblock, below=of dimensionreduction]  (network)     {Visualization};
    \path[draw,->] (data)      edge (preprocess1)
                   (preprocess1)      edge (topics)
                   (topics)      edge (subset)
                   (subset)    edge (corpus)
                   (corpus)   edge (preprocess2)
                   (preprocess2)       edge (parser)
                   (parser)   edge (extract)
                   (extract)    edge (triplets)
                   (triplets)    edge (dimensionreduction)
                   (dimensionreduction)  edge (network)
                    ;
    \end{tikzpicture}
    \caption{System pipeline}
\end{figure}

\subsection{Validations of topic modeling}
A key question for NMF topic modelling is to specify the number of topics K prior to running the algorithm. There is no consensus on a standardised approach to selecting the optimal number of topics. We chose to make use of the coherence value known as the C\_V measure proposed by \cite{Roderetal2015}. We ran 45 different NMF models using the values of K ranging from 5 to 50 (sequenced by interval of 1) and calculated the coherence value of each topic model. Figure A2 below shows the coherence scores of these 45 models. C\_V coherence score creates content vectors of words using their co-occurrences and calculates the score using normalised point-wise mutual information and the cosine similarity. C\_V takes a value between 0 and 1 and higher values indicate more coherent topic models. As Figure A2 has shown, 19 topics model has a reasonably high coherence score at 0.73 and interpretable topic results. We therefore chose 19 as the optimal number of topics and ran NMF. The list of topics that are identified as relevant events and their keywords with weights are as follows:

\textbf{Russia-Ukraine war}: [('USA', 16.22), ('Russia', 15.8), ('Ukraine', 12.23), ('State', 5.95), ('Sanction', 3.42), ('President', 2.66), ('Europe', 2.57), ('Western', 2.52), ('NATO', 2.46), ('War', 2.36), ('Putin', 2.32), ('Report', 2.12), ('Global', 1.94), ('International', 1.82), ('Company', 1.82), ('Biden', 1.78), ('World', 1.74), ('Economy', 1.63), ('Situation', 1.62), ('Conflict', 1.59), ('Provide', 1.58), ('Support', 1.39), ('Influence', 1.38), ('Russia Troops', 1.38), ('EU', 1.37), ('Region', 1.36), ('Gas', 1.36), ('News', 1.34), ('Bio', 1.31), ('Military affairs', 1.28)]

\textbf{COVID Shanghai Lockdown}: [('Shanghai', 24.6), ('Hospital', 3.11), ('Supply', 2.39), ('Nucleic acid', 1.67), ('Fight Pandemic', 1.64), ('Pandemic', 1.6), ('Life', 1.38), ('Community', 1.34), ('Portable building', 1.21), ('Housekeeping', 1.2), ('Company', 1.15), ('Shanghai City', 1.13), ('Announce', 1.12), ('City', 1.07), ('Patients', 1.04), ('Hope', 0.98), ('Nationwide', 0.97), ('Support', 0.95), ('Quarantine', 0.94), ('Pandemic control', 0.92), ('Asymptomatic', 0.91), ('Elderly', 0.91), ('Residents', 0.86), ('Vegetables', 0.78), ('Support from far', 0.77), ('Doctors', 0.76), ('Find', 0.75), ('Come on', 0.71), ('Medical team', 0.68), ('People', 0.67)]

\textbf{Winter Olympics}: [('Beijing', 27.27), ('Winter Olympics', 13.35), ('Winter Olympics Event', 12.51), ('Paralympic Games', 3.72), ('Ice and snow', 3.24), ('Come on', 2.66), ('Bing Dun Dun', 2.34), ('Opening', 2.32), ('Athletes', 2.3), ('Ski', 2.25), ('Opening ceremony', 2.1), ('Top athlete', 1.79), ('Match', 1.76), ('Program', 1.67), ('Sport', 1.44), ('News', 1.44), ('Team China', 1.37), ('Gold medal', 1.34), ('Expect', 1.26), ('Sports', 1.23), ('China Unicom', 1.16), ('Eileen Gu', 1.13), ('Torch', 1.1), ('World', 1.1), ('Olympic', 1.08), ('Fabulous', 1.01), ('Beijing city', 1.0), ('Closing ceremony', 0.99), ('Beginning of Spring', 0.97), ('Speed skating', 0.97)]

\begin{figure}[!ht]
    \centering
    \includegraphics[width=\textwidth]{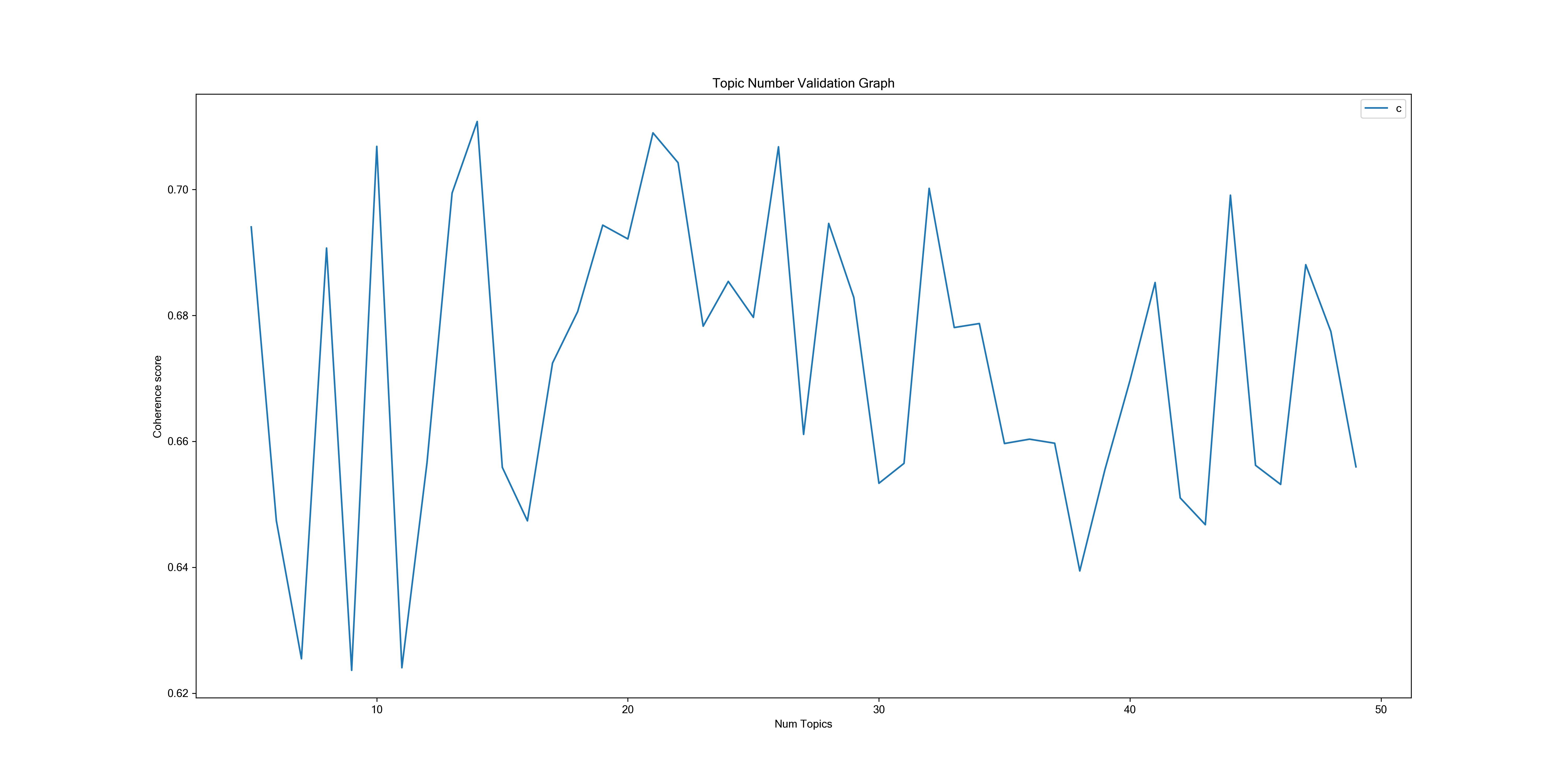}
    \caption{Validations of topic modeling}
\end{figure}

\begin{figure}[!ht]
    \centering
    \includegraphics[width=\textwidth]{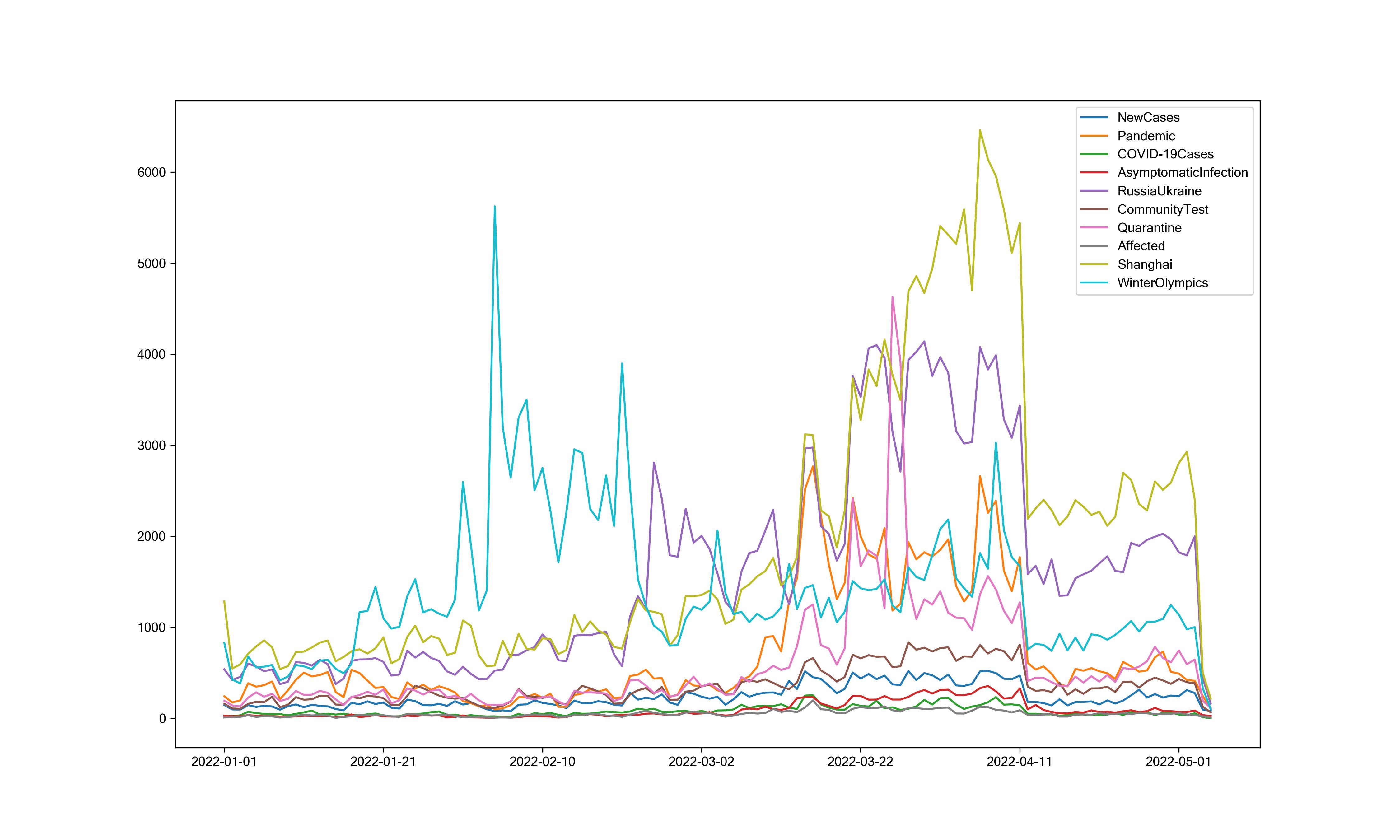}
    \caption{The change of selected key topics across time}
\end{figure}

\clearpage

\subsection{Visualization of four cases}

For easy interpretation and visualization, we select and compare the top 20 percent narratives from government/media accounts and celebrity accounts across the three events identified through topic modeling and all narratives on the Chained Woman Event.\footnote{For the Chained Woman event, given its sensitivity and the likelihood of distractions and censorship, there are fewer Weibo posts on this event. Therefore, mapping all narratives is necessary to capture the full range of narratives.} Figure A4 and A5 maps the top 20 percent narratives on the two international events: the Russia-Ukraine war and the Beijing Winter Olympics. Figure A6 maps the top 20 percent narratives on the Shanghai Lockdown during COVID-19 and Figure A7 presents all narratives on the Chained Woman Event. In these narrative figures, the nodes represent subjects and objects, while the arrows indicate directions from subjects to objects with verbs displayed as text on the arrows. In the network visualization,the thickness of the arrows represents the frequency of SVO triplets in the corpus, while node size indicates the frequency of actors appearing as either subjects or objects. All texts have been translated from Chinese to English by the authors.

\begin{figure}[p]
\centering
\subfloat[Government/Media accounts]{
    \includegraphics[width=0.6\textwidth]{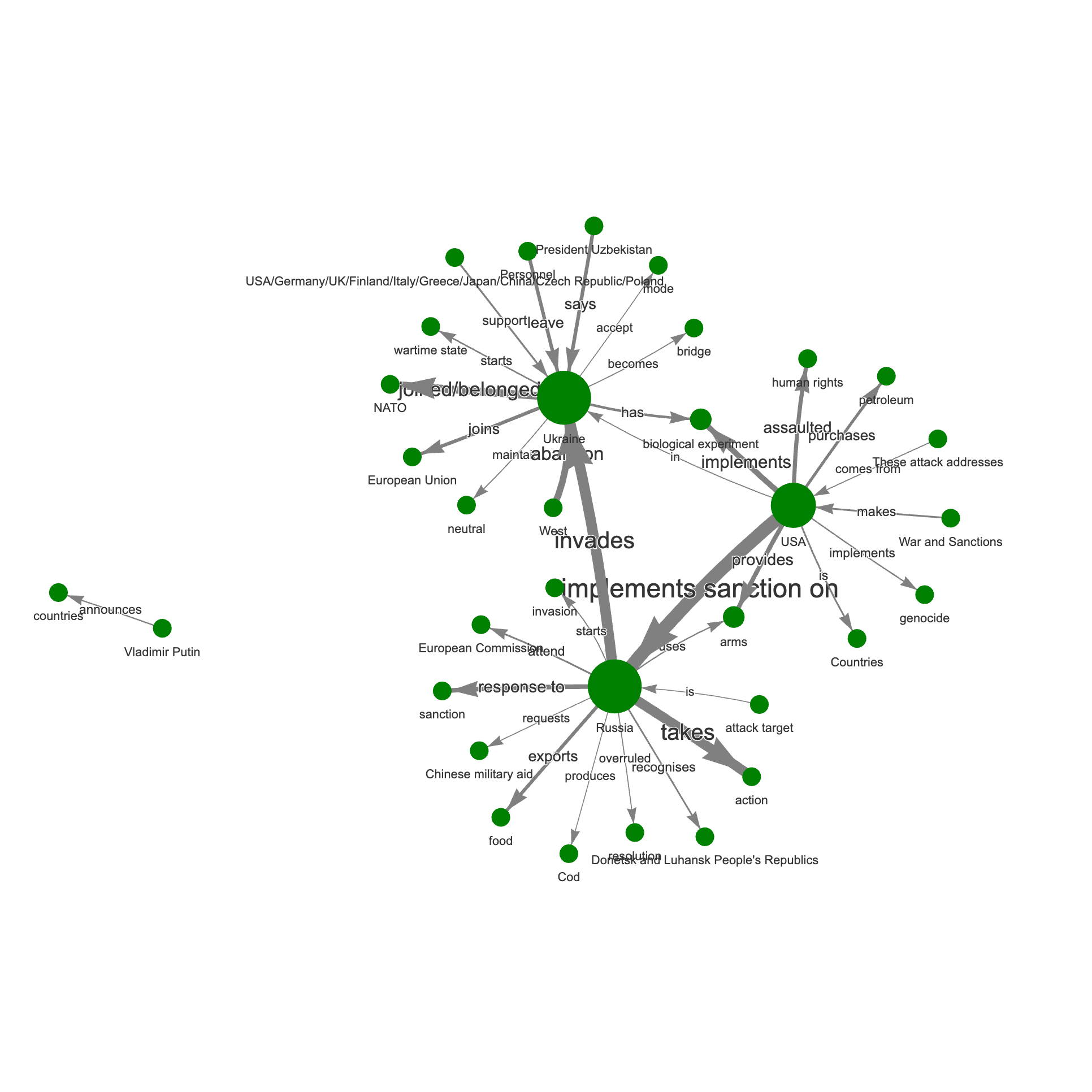}
}

\subfloat[Celebrity accounts]{
    \includegraphics[width=0.6\textwidth]{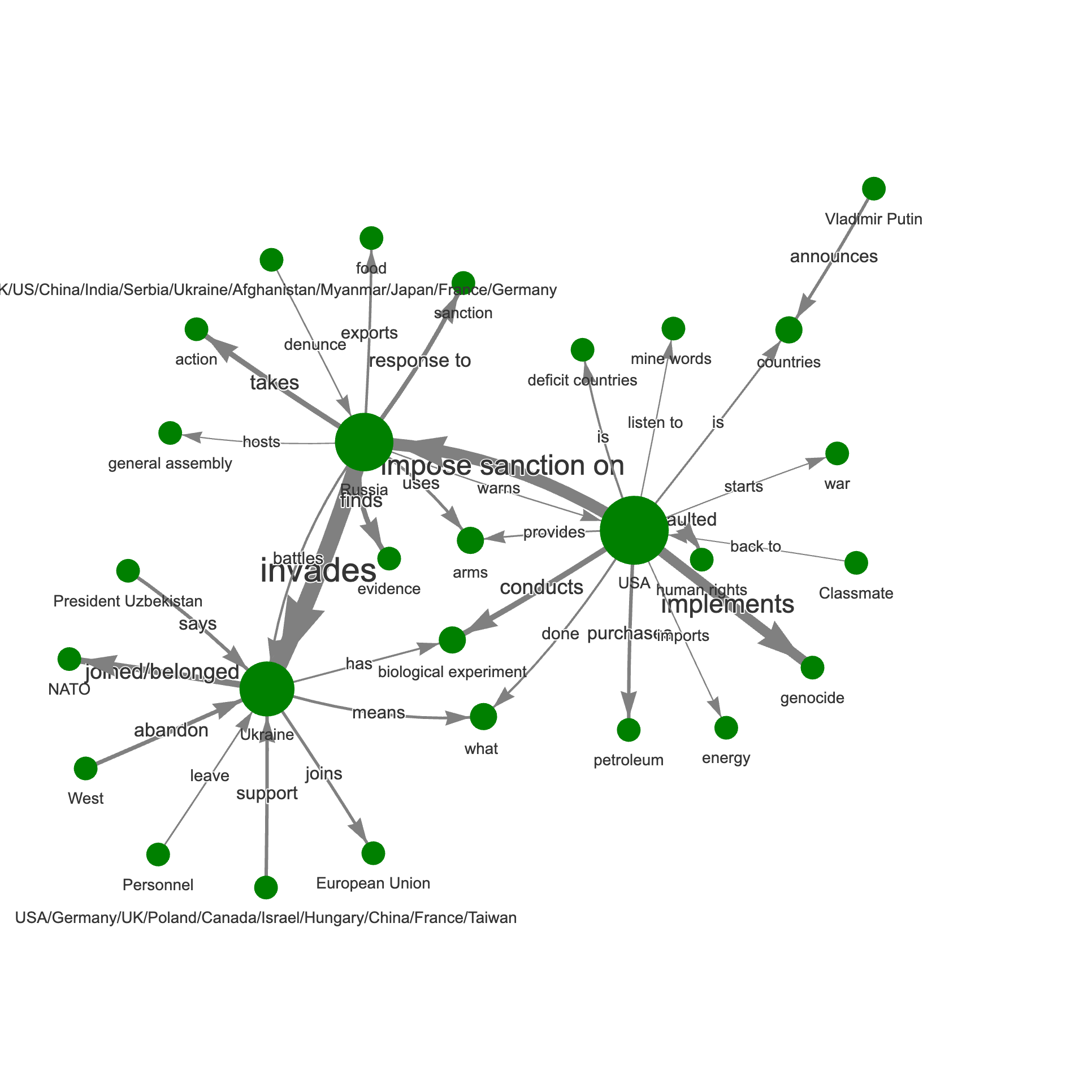}
}
\caption{Top 20\% narratives on Russia-Ukraine War by different accounts}
\end{figure}

\begin{figure}[p]
\centering
\subfloat[Government/Media accounts]{
    \includegraphics[width=0.6\textwidth]{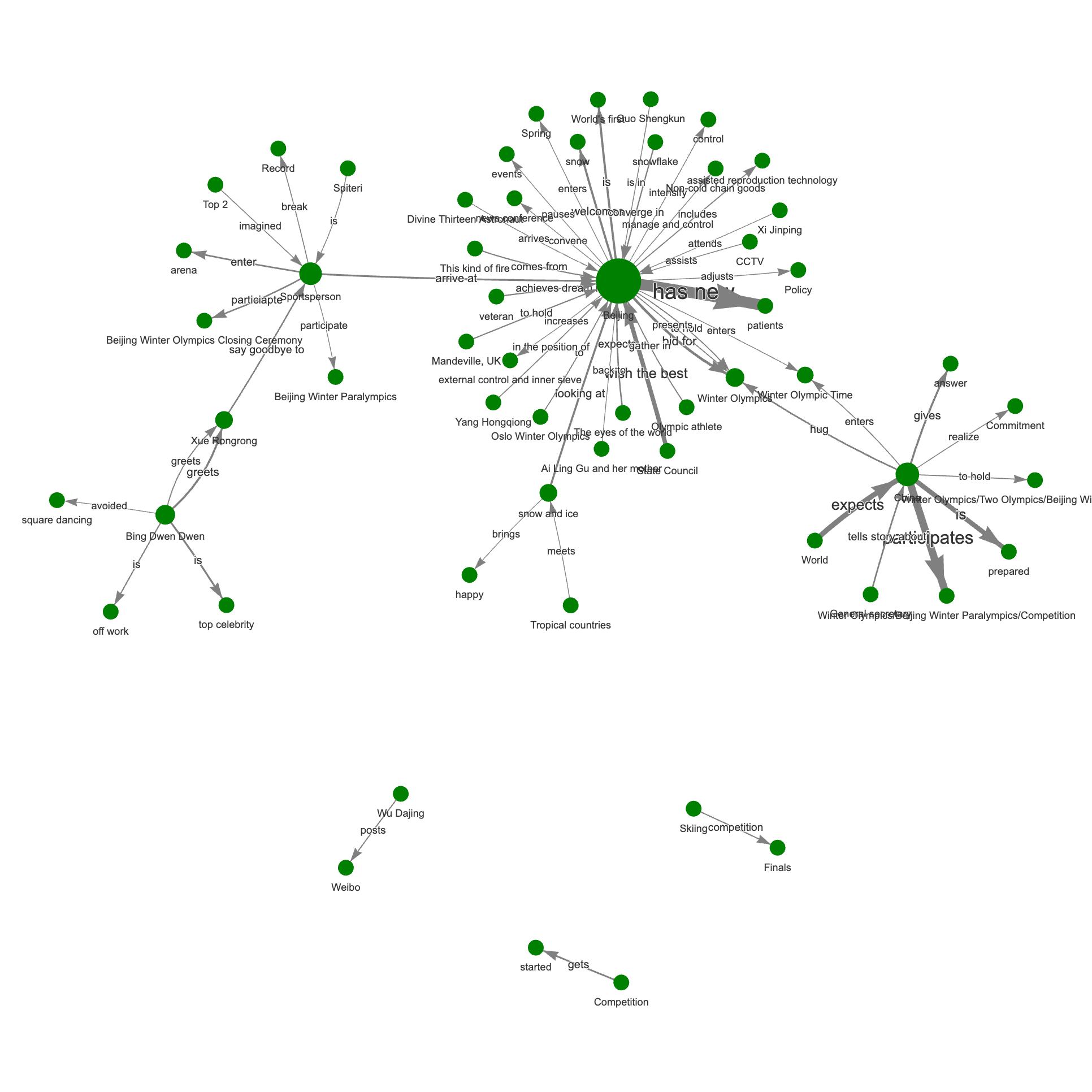}
}

\subfloat[Celebrity accounts]{
    \includegraphics[width=0.6\textwidth]{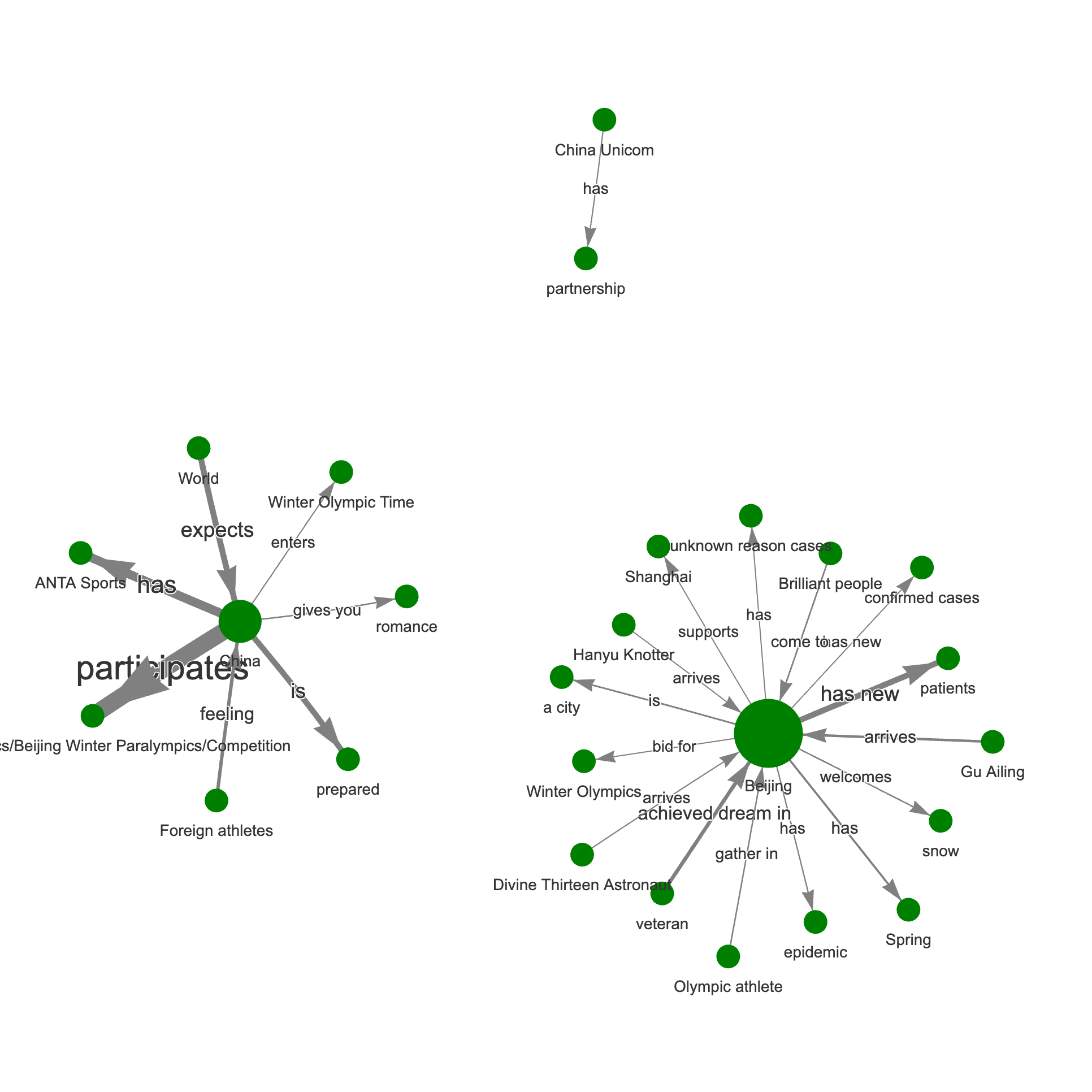}
}
\caption{Top 20\% narratives on Beijing Winter Olympics by different accounts}
\end{figure}

\begin{figure}[p]
\centering
\subfloat[Government/Media accounts]{
    \includegraphics[width=0.6\textwidth]{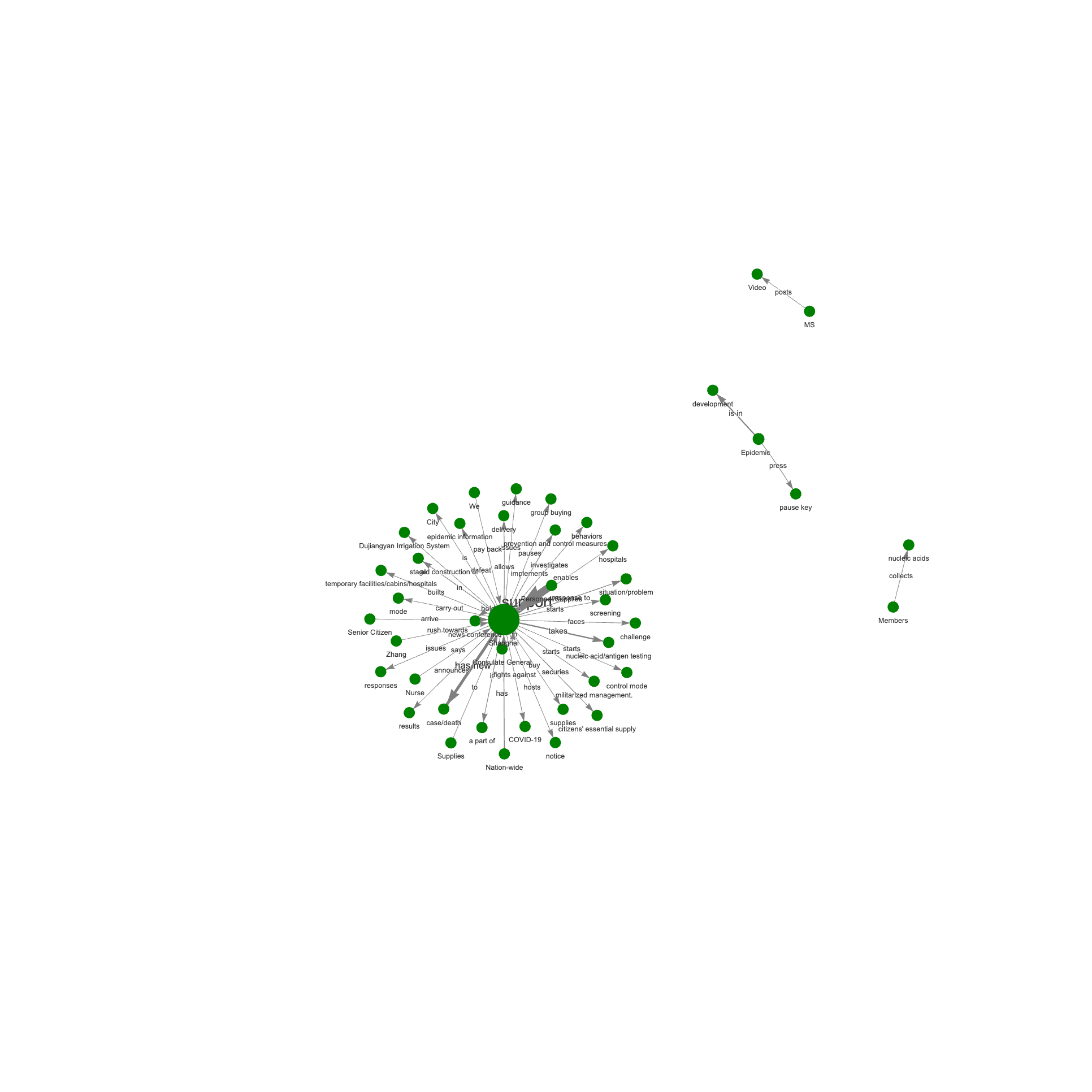}
}
\\[\dimexpr-15mm\relax]
\vspace{10mm}
\subfloat[Celebrity accounts]{
    \includegraphics[width=0.6\textwidth]{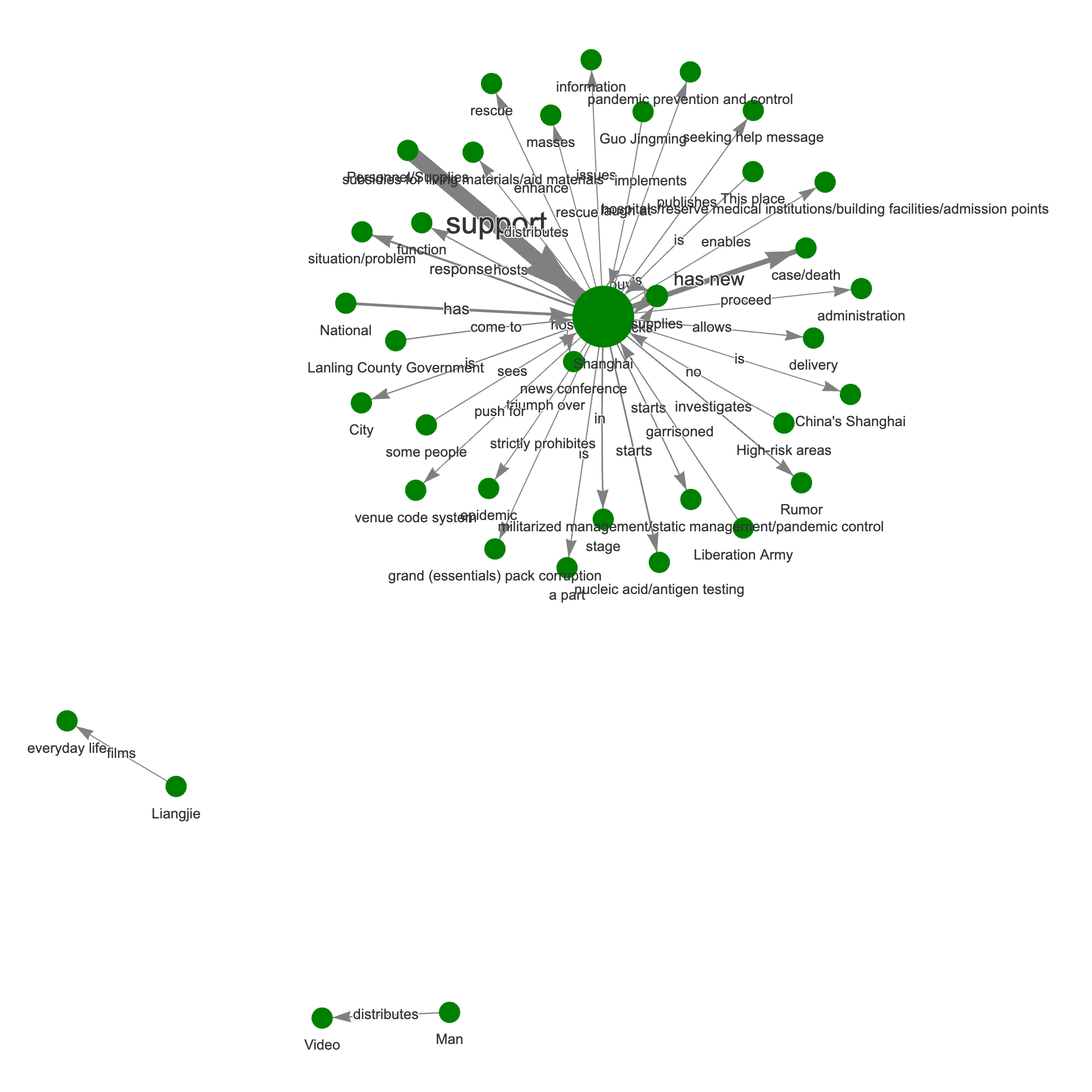}
}
\caption{Top 20\% narratives on Shanghai Lockdown by different accounts}
\end{figure}

\begin{figure}[p]
\centering
\subfloat[Government/Media accounts]{
    \includegraphics[width=0.6\textwidth]{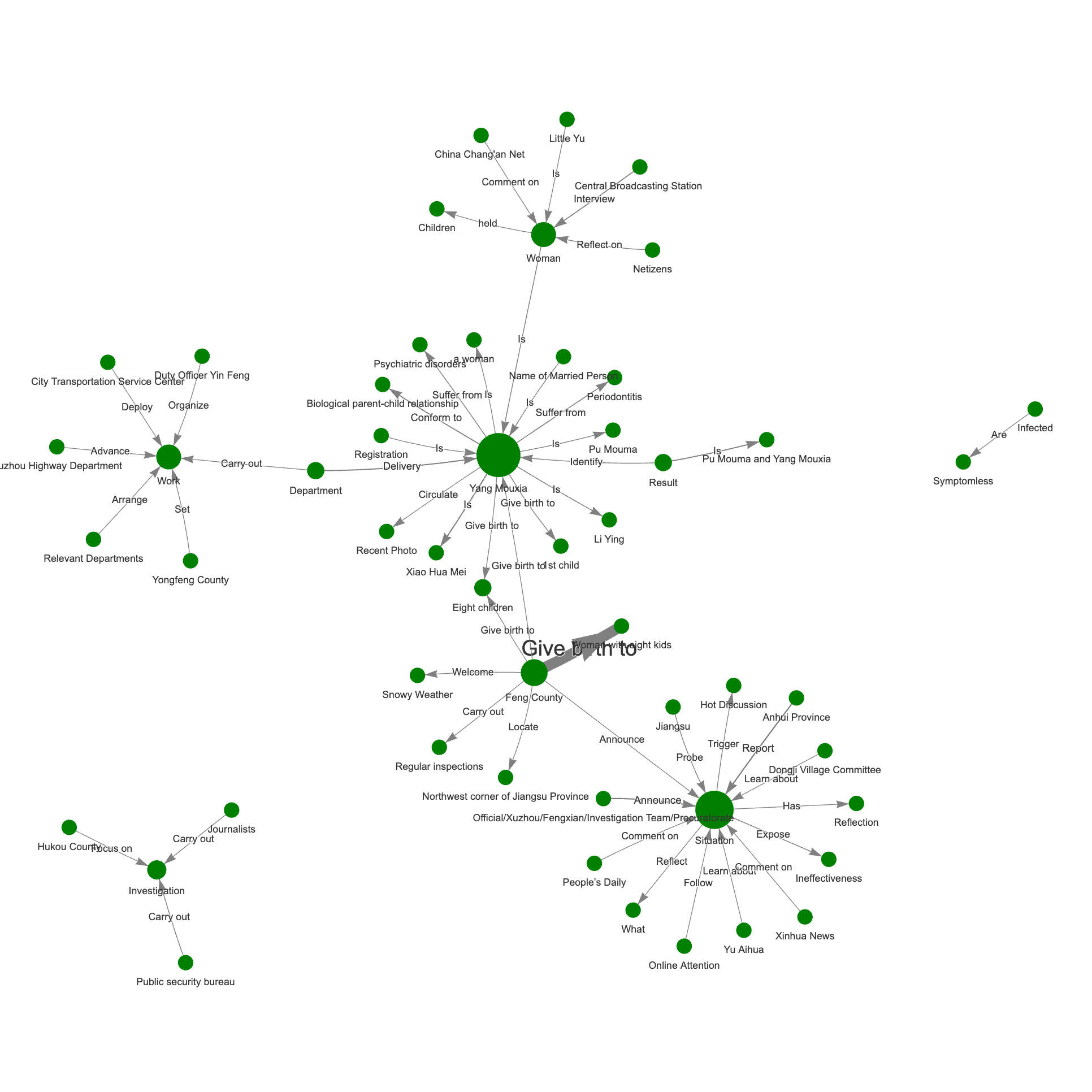}
}
\\[\dimexpr-15mm\relax]
\vspace{10mm}
\subfloat[Celebrity accounts]{
    \includegraphics[width=0.6\textwidth]{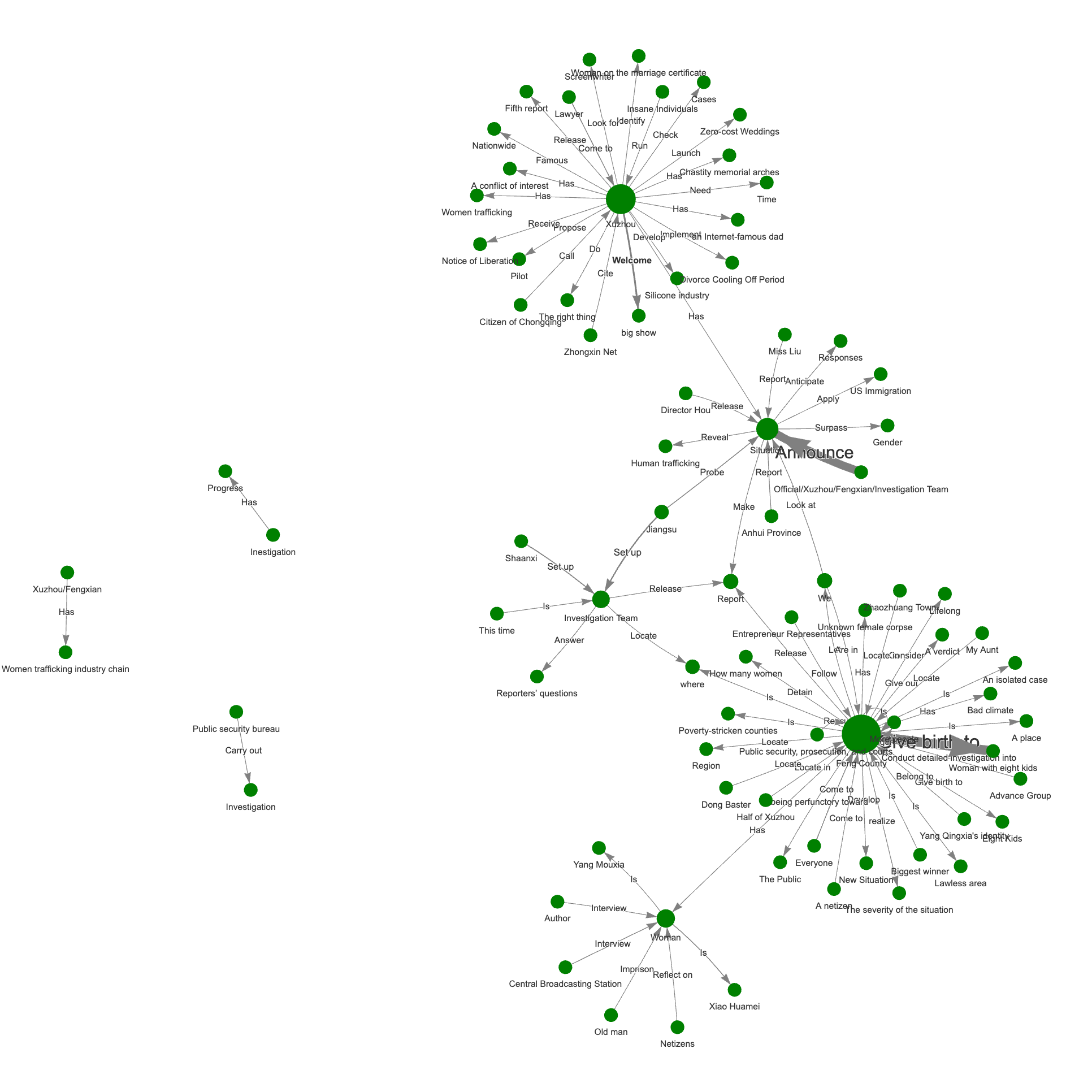}
}
\caption{All narratives on the Chained Woman event by different accounts}
\end{figure}

\clearpage

\section{Example quotes of key events}

\textbf{On Russia-Ukraine war by government and media accounts:}

[\#Putin says Ukraine joining NATO is completely unacceptable\# \#Putin decides to carry out a special military operation in Donbas\#] On the 24th local time, Russian President Vladimir Putin delivered an emergency televised address regarding the situation in Ukraine, stating that he has decided to launch a special military operation in the Donbas region. He also emphasized that Ukraine's accession to NATO is completely unacceptable.Russian President Putin has made the decision to conduct a special military operation in Donbas.Reporter: Zhang Yuyao, China Central Television News via Weibo video (China Central Television News, 2022-02-24, Attitude count: 319204, Comments count: 12741, Reposts count: 5932)

[Sanctions Have Extended to Russian Cats…] \#The West Begins Sanctioning Russian Civilians\# On March 1st, the Fédération Internationale Féline (FIFe) — the International Feline Federation — announced sanctions against Russia on its official social media account. The organization called on its national member associations to suspend the import of cats bred within Russia and ban organizations with ties to Russia from participating in FIFe-organized exhibitions and competitions.\#Russian Cats Sanctioned\#\#Western Sanctions Target Russian Civilians\# (Global Times, 2022-03-02, Attitude count: 2110287, Comments count: 64653, Reposts count: 61396)

[\#Ukrainian President Says the West Has Completely Abandoned Ukraine\#] On the 25th local time, Ukrainian President Volodymyr Zelensky delivered a video address, stating that the West has completely abandoned Ukraine.
“I asked the leaders of 27 European countries: Will Ukraine join NATO? But everyone was afraid. No one answered me,'' he said. “We are not afraid to talk with Russia. We are ready to discuss all issues — including our country's security guarantees and neutral status.'' People’s Daily, via CCTV News Weibo video. (Zhenjiang Government News Release, 2022-02-25, Attitude count: 2, Comments count: 0, Reposts count: 1)

[\#Understand the Root Causes of the Ukraine Crisis in 5 Minutes\#] According to China Central Television (CCTV), the situation in Ukraine has recently escalated sharply, with ongoing clashes between Russian and Ukrainian forces. Some analysts believe that the root cause of the Ukraine crisis lies in the continuous expansion of NATO, led by the United States, which has gradually moved closer to Russia's borders, ultimately impacting Russia’s core national security interests.
Watch the video to understand more \#Analysts Say the Root Cause of the Ukraine Crisis Is NATO's Eastward Expansion Threatening Russian Security\#
Source: People's Daily, via Weibo video. (Southern Daily, 2022-02-28, Attitude count: 18, Comments count: 3, Reposts count: 4)

[Ministry of Foreign Affairs: \#The U.S. Is the Root Cause of the Escalating Ukraine Crisis\#] Today, China's Ministry of Foreign Affairs once again addressed the Ukraine issue. Spokesperson Hua Chunying stated:
“The key question now is: What role has the U.S., the initiator of the current tense situation in Ukraine, played in this crisis? What impact has it had?” She added: “If someone keeps adding fuel to the fire while blaming others for not putting it out, such behavior is both irresponsible and immoral.” Source: CCTV Military, via Weibo video. (Lanzhou News Release, 2022-02-24, Attitude count: 0, Comments count: 0, Reposts count: 0)

\bigskip
\textbf{On Russia-Ukraine war by celebrity accounts:}

U.S. Secretary of State Antony Blinken and Secretary of Defense Lloyd Austin visited Ukraine, where the devastation of war paved a “red carpet” for them—the true masterminds behind the conflict.Although the war is being fought between Russia and Ukraine, neither country truly wanted it. Both are being weakened by the ongoing conflict. Europe, too, has become a loser in this war, while the only real beneficiary is the United States.
In this sense, Blinken and Austin came to “inspect” the war, to see how it is serving U.S. strategic interests.I believe that no one in today’s world truly wants to fight a war. Most wars arise when parties feel they have no better option—believing that fighting may bring a better outcome than not fighting.However, what makes the Russia-Ukraine war so unpredictable and difficult to calculate is the presence of a powerful third party—the United States—whose stance has a profound impact on the direction and outcome of the war.Right now, the U.S. is profiting, Ukraine is in ruins, and Russia is wounded. The United States has shed no blood, printed a bit more money, and gained greater control over the situation—tightening its grip on Europe in the process.America remains incredibly powerful—but also quite devious.It’s crucial not to fight for America’s prosperity, nor to make sacrifices while harboring anti-American sentiment—only to let the U.S. emerge as the ultimate winner.This is a strategic equation—a formula of gain and loss—that all regions experiencing geopolitical tension must keep in mind.\#V-Insight Commentary\# (Guest Commentator for Global Times, 2022-04-26, Attitude count: 16769, Comments count:1991 , Reposts count:378)

As long as we're alive, that's enough.\#Ukraine\#The terrifying Ukrainian Nazis are abusing captured Russian soldiers.(Science Content Creator, 2022-03-03, Attitude count: 1804, Comments count:230 , Reposts count:223)

Repost[\#Biden Delivers Remarks on U.S. Security Assistance to Ukraine\#]On May 3rd local time, U.S. President Joe Biden delivered a speech on U.S. security assistance to Ukraine during a visit to a weapons manufacturing facility in Troy, Alabama.He stated that the United States, in coordination with its allies and partners, is providing strong support to Ukraine. Since the start of the Russia-Ukraine conflict, the U.S. alone has provided over \$3 billion in security assistance to Ukraine.Source: CCTV Military, via Weibo video.\#RussiaUkraineConflictUpdate (Top Super Topic Fan, 2022-05-04, Attitude count: 0, Comments count:0 , Reposts count:0)

\bigskip
\textbf{On Beijing Winter Olympics by government and media accounts:}

[\#The Seeds Planted by the Beijing Winter Olympics Have Been Sown Across the World\#] Over the past few weeks, the Beijing Winter Olympics has left us with countless unforgettable moments.From the stunning, continuous-shot displays of Chinese cultural beauty in the opening and closing ceremonies, to the passion and perseverance of athletes on the field, and the touching acts of mutual encouragement and human warmth off the field—these memories have come together to form something truly eternal.The values conveyed and the seeds planted have now been sown across the world. We look forward to seeing these seeds take root, grow, and blossom—like willows turning into brushes, writing a future that reaches beyond borders.Let’s work together to create a brighter tomorrow!

\#AnchorCommentary from Xinwen Lianbo\# Source: CCTV News via Weibo Video. (CCTV News, 2022-02-21, Attitude count: 2212, Comments count:246, Reposts count:309)

\#Winter Olympics Theme MV — A Promise to the World\#On February 4, the world will turn its eyes to the opening of the Beijing 2022 Winter Olympics.In celebration, Xinhua News Agency has released an original theme music video titled “A Promise to the World'', featuring Chinese singer Li Yuchun (Chris Lee).“The world is looking to China, and China is ready.”A grand winter gathering, a springtime invitation —Let the power of music cheer on the Beijing Winter Olympics!Source: Xinhua News Agency, via Weibo video (Ezhou Government Website, 2022-02-04, Attitude count: 71, Comments count:14, Reposts count:29)

\#Feels Like the Opening Ceremony Was Just Yesterday\#[This Moment Was Electrifying! \#National Flag Raised and Anthem Played at Winter Olympics Closing Ceremony\#]I love my motherland, and I’m proud of China!\#BeijingWinterOlympicsClosingCeremony.Source: CCTV News via Weibo Video. (CCTV News, 2022-02-20, Attitude count: 60617, Comments count:2126, Reposts count:5744)

\bigskip
\textbf{On Beijing Winter Olympics by celebrity accounts:}

What was the biggest difference between the 2022 Winter Olympics opening ceremony and that of 2008? Zhang Yimou, director of both events, said something that moved many people:“Chinese people are just like everyone else—so sincere, so kind, so in love with beauty, so romantic. I hope everyone in the world can be well.”\#TheBiggestDifferencein2022BeijingWinterOlympics\# (An influencer, 2022-02-04, Attitude count: 280645, Comments count:8619, Reposts count:111060)

@sy Wang Shiyue @Ice-dance Liu Xinyu Congratulations on your wonderful performance at the Winter Olympics! You were amazing — so proud of you both! Thank you so much for your support, I'm truly honored! Looking forward to your gala performance on the 20th — see you on the ice! (A celebrity, 2022-02-17, Attitude count: 4729877, Comments count:1000000, Reposts count:1000000)

\#Every Shot at the Winter Olympics Has ANTA\#
Netizens' focus this year has really taken some unexpected turns — hilarious! At the Beijing Winter Olympics, ANTA seems to be saying:“If there's even one camera angle without our logo, I’ll be heartbroken, OK?” Gotta admit — Eileen Gu’s golden dragon suit is super stylish, keeps her warm and makes her look slim. And the ink-splash design on the curling team uniforms totally radiates that traditional Chinese vibe. I hereby declare that domestic Chinese brands have officially won my heart! (An influencer, 2022-02-17, Attitude count: 11363, Comments count:2123, Reposts count:1626)

\bigskip
\textbf{On Shanghai lockdown by government and media accounts:}

Without a people's army, there is nothing for the people. \#People's Army Supporting Shanghai\# (The Central Committee of the Chinese Communist Party's Youth League, 2022-04-04, Attitude count: 13676, Comments count: 848, Reposts count: 1604)

[Jin Yin Tan Head Nurse: \#Shanghai is a part of our Jin Yin Tan\#] 'Because one person falls in love with a city, and because Shanghai has Director Zhong Ming and the Shanghai medical team, I feel that Shanghai is a part of Jin Yin Tan.' During the Wuhan epidemic, Cheng Fang served as the ICU head nurse at Jin Yin Tan Hospital. She always remembers that the Shanghai medical aid team was the first medical team to enter Jin Yin Tan. Now she is supporting nucleic acid sampling in Xinjing, Changning District. Weibo video from SMGNEWS (China Daily, 2022-04-05, Attitude count: 96263, Comments count: 5892, Reposts count: 2004)

[\#Shanghai Responds to Claims About Yunnan Supplies\# Supplies Being Sorted for Distribution to Workers] According to Shanghai Network Rumor Refutation: On April 30, a netizen posted on Weibo claiming that at a construction site near Chujia Road in Qingcun Town, Fengxian District, Shanghai, “supplies from Yunnan are unwanted, some have spoiled, and they're being dumped at construction sites.'' From the accompanying video, vegetables including potatoes, peppers, onions, and eggplants were taken out of supply boxes and piled together, with no obvious signs of spoilage. In response, @FengxianQingcun issued a statement expressing serious concern about the situation reported on Weibo. After investigation, they clarified that the scene shown in the video was actually workers at the construction site sorting and bagging fresh vegetables for distribution, not “unwanted supplies'' or “spoiled goods dumped at the construction site'' as the blogger claimed. Qingcun Town has reported the netizen to the police for deliberately distorting facts, spreading false information, confusing public opinion, and misleading the public. (Suihua Procuratorate, 202-05-01, Attitude count: 0, Comments count: 0, Reposts count: 0)

\bigskip
\textbf{On Shanghai lockdown by celebrity accounts:}

\#celebrity nameDonatesToShanghaiEpidemicPrevention\# \#celebrity name\# donated epidemic prevention supplies to Shanghai, focusing on specialized communities that don't usually receive much attention. He has always done more than he's said, and he is our forever kind-hearted \#celebrity name\# who continues to spread positive energy. (Well-known entertainment blogger, 2022-04-20, Attitude count: 112755, Comments count: 6685, Reposts count: 8024)

Curse word! //@XX: I once imagined that North Town would someday be known throughout the country. Perhaps for its pear blossoms that bloom across the city in April, perhaps for the rich cultural heritage of the thousand-year-old ancient city, perhaps for the Medicine Shaman North God Mountain that forever guards the frontier, or perhaps for its honest people, delicious food, or even Manchu culture... But it should not become famous because vegetables sent to aid epidemic areas were thrown into garbage dumps. After learning that a certain area was “short of supplies,'' a massive volunteer team of over 5,000 people—composed of Communist Party members, teachers, and grassroots officials—gathered from all over the city at 2 AM to harvest vegetables. Later, more than 1,600 volunteers formed teams to sort and package these plump cabbages, potatoes, and carrots, loading them onto 42 large trucks that rushed to provide aid through the night. This manpower and these supplies nearly exhausted the entire resources of this small county. These supplies, which could be stored for a whole winter in the frontier region, were escorted by police officers and sent with the expectations of the entire city's people. After going through cold chain logistics and ferry transportation, they arrived at the place experiencing supply shortages in just one day.Just as we were eagerly waiting for these supplies to alleviate the severe pressure on resources, news came from the distant riverside: they were thrown into garbage dumps due to “spoilage during transportation.'' There's no point saying more—it's not just the vegetables that have rotted, but perhaps also the hearts of certain people.(An influencer and photographer, 2022-04-19, Attitude count: 0, Comments count: 0, Reposts count: 0)

\bigskip
\textbf{On the Chained Woman event by government and media accounts:}

[\#Jiangsu Provincial Party Committee and Government Investigation Team Announces\#\#Investigation and Handling of the Feng County Woman Who Gave Birth to Eight Children Case\#] Announcement from the Jiangsu Provincial Party Committee and Government Investigation Team regarding the investigation and handling of the 'Feng County Woman Who Gave Birth to Eight Children' incident'. (The News Center of Chinese Central Television Station, 2022-02-23, Attitude count: 2,487,313, Comments count: 147,699, Reposts count: 269,465).

[\#Investigation Team Established for the Feng County Woman Who Gave Birth to Eight Children Incident\#] The Jiangsu Provincial Party Committee and Provincial Government have decided to establish an investigation team to thoroughly investigate the 'Feng County woman who gave birth to eight children' incident, comprehensively uncover the truth, severely punish relevant illegal criminal behaviors according to law, hold responsible personnel strictly accountable, and publish the results to the public in a timely manner. (The News Center of Chinese Central Television Station, 2022-02-23, Attitude count: 1,473,092, Comments count: 167,200, Reposts count: 606,327).

Forwarded Weibo: [\#Complete Investigation of the Feng County Woman Who Gave Birth to Eight Children Incident\#] From her appearance in Dongji Village in 1998 to the exposure of the incident in January 2022, spanning 24 years, why wasn't Xiaohuamei's tragedy discovered and stopped in time? \#Media Reveals Investigation Process of Woman Who Gave Birth to Eight Children Incident\# Investigation of the 'Feng County woman who gave birth to eight children' incident. (The official Weibo account of a county level media center, 2022-02-23, Attitude count:0, Comments count: 0, Reposts count: 0)

\bigskip
\textbf{On the Chained Woman event by celebrity accounts:}

Published a blog post 'The Xuzhou February 7th Investigation Report on the “Chained Woman'' Contains Reasoning Errors, Evades Key Issues, and Its Conclusions Cannot Stand' Note: I had planned to review case files this morning. After waking up and seeing the report released by Xuzhou, I was extremely angry! What kind of quality is this?! As a criminal defense lawyer who routinely authenticates and argues evidence in dozens or hundreds of case files, [I find that] the Xuzhou February 7th investigation report on the “Chained Woman'' contains reasoning errors, evades key issues, and its conclusions cannot stand  (Partner at a law firm, 2022-02-08, Attitude count: 98,524, Comments count: 2,528, Reposts count: 54,649)

//@XXXX: Everyone, repost this until it reaches 200,000 reposts to achieve a Level 1 public opinion event. Don't scatter the reposts. We hope that the Central Government will directly dispatch an investigation team to investigate the related incidents in Feng County, Xuzhou, Jiangsu Province, and thoroughly investigate whether there is a human trafficking industry chain and protective umbrella in the local area.(An original video blogger/Entertainment blogger, 2022-02-12, Attitude count: 0, Comments count: 0, Reposts count: 1)

\#Investigation and Handling of the Feng County Woman Who Gave Birth to Eight Children Case\# Is there a big difference from before? It's just that some people received internal Party disciplinary actions. (Super Topic Fan Influencer, 2022-02-23, Attitude count: 0, Comments count: 0, Reposts count: 0)

\end{appendix}

\end{document}